\documentclass[10pt,journal,compsoc]{IEEEtran}

\usepackage{epsfig,endnotes}

\usepackage{pdflscape}

\usepackage{lscape, latexsym, amssymb, amsmath, multirow}
\usepackage[linesnumbered, vlined, ruled]{algorithm2e}
\usepackage{graphicx}
\usepackage{subfigure}
\usepackage{multirow}
\usepackage{mathtools, bbm, color}
\usepackage{pdfpages}
\usepackage{rotating}
\usepackage{amsmath,amssymb,amsthm,mathrsfs,amsfonts,dsfont}

\usepackage{pifont, textcomp, marvosym, wasysym, extarrows}
\usepackage{tikz}

\ifCLASSOPTIONcompsoc
  \usepackage[nocompress]{cite}
\else
  \usepackage{cite}
\fi
\ifCLASSINFOpdf
\else
\fi
\begin{document}
%
\title{Adversarial CAPTCHAs}

\author{Chenghui~Shi,
	Xiaogang~Xu,
	Shouling~Ji,
	Kai~Bu,
	Jianhai~Chen,
	Raheem~Beyah,
	and Ting~Wang
	\IEEEcompsocitemizethanks{
		
%
%
%
		\IEEEcompsocthanksitem C. Shi, X. Xu, K. Bu and J. Chen are with the College of Computer
		Science and Technology, Zhejiang University, Hangzhou, Zhejiang 310027, China.\protect\\
		E-mail: \{chenghuishi, xiaogangxu, kaibu, chenjh919\}@zju.edu.cn
		\IEEEcompsocthanksitem S. Ji is with the College of Computer Science and Technology, Zhejiang University, Hangzhou, Zhejiang 310027, China, and with the School of Electrical and Computer Engineering, Georgia Institute of Technology, Atlanta, GA 30332, USA. \protect\\
		E-mail: sji@zju.edu.cn
		\IEEEcompsocthanksitem R. Beyah is with the School of Electrical and Computer
		Engineering, Georgia Institute of Technology, Atlanta, GA 30332, USA. \protect\\
		E-mail: rbeyah@ece.gatech.edu
		\IEEEcompsocthanksitem T. Wang is with the Department of Computer Science and Engineering,
		Lehigh University, Bethlehem, PA 18015, USA. \protect\\
		E-mail: inbox.ting@gmail.com
	}
}%

\IEEEtitleabstractindextext{%
	\begin{abstract}
		Following the principle of \emph{to set one's own spear against one's own shield},
		we study how to design adversarial CAPTCHAs in this paper.
		We first identify the similarity and difference between adversarial CAPTCHA generation
		and existing hot adversarial example (image) generation research.
		Then, we propose a framework for text-based and
		image-based adversarial CAPTCHA generation on top of
		state-of-the-art adversarial image generation techniques.
		Finally, we design and implement an adversarial CAPTCHA generation and evaluation system,
		named aCAPTCHA, which integrates 10 image preprocessing techniques,
		9 CAPTCHA attacks, 4 baseline adversarial CAPTCHA generation methods,
		and 8 new adversarial CAPTCHA generation methods.
		To examine the performance of aCAPTCHA, extensive security and usability
		evaluations are conducted. The results demonstrate that
		the generated adversarial CAPTCHAs can significantly
		improve the security of normal CAPTCHAs while maintaining
		similar usability.
		To facilitate the CAPTCHA security research, we also open source
		the aCAPTCHA system, including the source code, trained models,
		datasets, and the usability evaluation interfaces.
	\end{abstract}
	
	}

\maketitle

\IEEEdisplaynontitleabstractindextext

%
\IEEEpeerreviewmaketitle

\ifCLASSOPTIONcompsoc
\IEEEraisesectionheading{\section{Introduction}\label{sec:introduction}}
\else
\section{Introduction}
\label{sec:introduction}
\fi

%
%
%
%
%
%
CAPTCHA (Completely Automated Public Turing test to tell Computers and Humans Apart)
is a type of \emph{challenge-response test} in computing
which is used to distinguish between human and automated programs (machines).
The first generation of CAPTCHA was invented in 1997, while the term ``CAPTCHA"
was first coined in 2002 \cite{captchawiki}\cite{captchanet}.
Ever since its invention, CAPTCHA has been widely used to improve the security
of websites and various online applications to prevent the abuse of online services,
such as preventing phishing, bots, spam, and Sybil attacks.

\vspace{2mm}

\textbf{Existing CAPTCHA Schemes.}
In general, existing popular CAPTCHAs can be classified into four
categories:

(1) \emph{Text-based CAPTCHA}. Text-based CAPTCHA schemes
ask users to recognize a string of distorted
characters with/without an obfuscated background \cite{burmarccs11}\cite{gaoyanndss16}.
Due to its simplicity and high efficiency, text-based CAPTCHA
is the most widely deployed and acceptable form up to now
and in a foreseeable future \cite{burmarccs11}\cite{gaoyanndss16}.

(2) \emph{Image-based CAPTCHA}. Image-based CAPTCHA is another popular scheme
which usually asks users to select one or more
images with specific semantic meanings from a couple of candidate images \cite{chetygicis04}.
It is motivated by the intuition that compared with a string of characters,
images carry much richer information and have a larger variation space.
Meanwhile, there are still many hard, open problems in image perception
and interpretation, especially in the context of noisy environments.
Thus, to some extent, image-based CAPTCHA is more secure than text-based CAPTCHA.
Nevertheless, to the best of our knowledge, a comprehensive comparative
analysis on the security and usability of text- and image-based CAPTCHAs
is still void. Recently, many variants of image-based CAPTCHAs were proposed,
such as \emph{slide-based CAPTCHA} which asks users to slide a puzzle to the right part
of an image \cite{alealepatent12}, \emph{click-based CAPTCHA} which asks users to click specific semantic
regions of an image \cite{hwahuaisbast12}, etc.

(3) \emph{Audio-based CAPTCHA}.
Audio-based CAPTCHA
asks users to recognize the voice contents in a piece of audio
\cite{captchawiki}\cite{captchanet}. In most of the practical applications,
audio-based CAPTCHA
is often used together with text-based CAPTCHA as a complementary means,
mainly because of the usability issue, especially for non-native users
of the audio language.

(4) \emph{Video-based CAPTCHA}. Video-based CAPTCHA is a new kind of CAPTCHA
that asks users to finish a content-based video labeling task \cite{kluzansoups09}.
It is usually more complex and takes more time for users to correctly
finish compared with other forms of CAPTCHAs. Thus, it is not widely adopted and
seldom to see in practice.

There are also other different proposals for CAPTCHA design, e.g., game-based CAPTCHA \cite{xureyusenixsecurity12} and inference-based CAPTCHA \cite{sahnagitp15}.
However, they are not widely deployed yet due to various reasons, e.g., security issues,
accessability limitations, and performance issues.
In this paper, our study mainly focus on \emph{text- and image-based CAPTCHAs}.
The reason is evident: they are the most accepted and widely used CAPTCHAs
up to now and in a foreseeable future. The study of their security and usability
has more potential implications for practical applicaitons.

\vspace{2mm}
\textbf{Issues of CAPTCHAs and Motivation.}
Generally speaking, CAPTCHA can be evaluated according to its \emph{security performance},
which refers to the strength and resilience of CAPTCHAs against various attacks,
and \emph{usability performance}, which refers to how user friendly the CAPTCHAs are
\cite{captchawiki}\cite{captchanet}.
From the security perspective, it is not a news to see reports that a CAPTCHA scheme
is broken by some attacks \cite{captchawiki}\cite{captchanet}.
The evolution of CAPTCHAs always moves forward in a spiral, constantly accompanied by
emerging attacks.
For text-based CAPTCHAs, the security goal of its earliest version is to defend
against Optical Character Recognition (OCR) based attacks.
Therefore, many distortion techniques (e.g., varied fonts, varied font sizes, and rotation)
are applied. Over the last decade, machine learning algorithms become
more and more powerful. Following the seminal work which demonstrates that computers turn
to outperform humans in recognizing characters, even under severe distortion,
many successful attacks to text-based CAPTCHAs were proposed, including both
generic attacks which target multiple text-based CAPTCHAs \cite{burmarccs11}\cite{gaoyanndss16},
and specialized attacks which targeted one kind of text-based CAPTCHAs \cite{gaowanccs13}.
In spite of that it is possible to improve the security of text-based CAPTCHAs
by increasing the distortion and obfuscation levels, their usability will be
significantly affected \cite{burmarccs11}\cite{gaoyanndss16}.

The same dilemma exists for image-based CAPTCHAs either.
With the prosperity of machine learning research, especially
recent deep learning progress, Deep Neural Networks (DNNs) have
achieved impressive success in image classification/recognization,
matching or even outperforming the cognitive ability of humans
in complex tasks with thousands of classes \cite{hezhacv15}.
Along with such progress, many DNN-based attacks have been proposed recently to
crack image-based CAPTCHAs with very high success probability, as demonstrated
by a large number of reports \cite{sivpoleurosp16}.
To defend against existing attacks, the intuition is to rely on high-level
image semantics and develop more complex image-based CAPTCHAs,
e.g., recognizing an image object by utilizing its surrounding context \cite{zhuyanccs10}.
Leaving the security gains aside, such designs usually induce poor usability \cite{captchawiki}\cite{captchanet}.
To make things worse, unlike text-based CAPTCHAs,
it is difficult, if not impossible, for designers to generate
specific images with required semantical meanings through certain rules.
In other words, it is too labor-intensive to collect labeled images
in large scale.

In summary, existing text- and image-based CAPTCHAs are facing challenges
from both the security and the usability perspectives.
It is desired to develop a new CAPTCHA scheme that achieves high security
while preserving proper usability, i.e., seeks a better balance between
security and usability.

\vspace{2mm}
\textbf{Our Methodology and Contributions.}
To address the dilemma of existing text- and image-based CAPTCHAs,
we start from analyzing state-of-the-art attacks.
It is not surprising that most, if not all, of the attacks to
text- and image-based CAPTCHAs
are based on machine learning techniques, especially the latest
and most powerful ones, which are mainly based on deep learning, typically, CNNs.
This is mainly because the development of CAPTCHA attacks roots in
the progress of machine learning research, as we discussed before.

On the other hand, with the progress of machine learning research,
researchers found that many machine learning models, especially neural networks,
are vulnerable to \emph{adversarial examples}, which are defined as elaborately
(maliciously, from the model's perspective) crafted inputs that are imperceptible
to humans but that can fool the machine learning model into producing
undesirable behavior, e.g., producing incorrect outputs \cite{szezararxiv13}.
Inspired by this fact, is that possible for us to design a new kind of CAPTCHAs
by proactively attacking existing CAPTCHA attacks,
i.e., ``\emph{to set one's own spear against one's own shield}"?

Following this inspiration, we study the method to generate text-
and image-based CAPTCHAs based on adversarial learning,
i.e., \emph{ text-based adversarial CAPTCHAs} and
\emph{ image-based adversarial CAPTCHAs}, that are resilient to state-of-the-art
CAPTCHA attacks and meanwhile preserve high usability.
Specifically, we have three main objectives in the design:
(1) \emph{security}, which implies that the developed CAPTCHAs can effectively
defend against state-of-the-art attacks, especially the powerful deep learning
based attacks; (2) \emph{usability}, which implies that the developed CAPTCHAs
should be usable in practice and maintain high user experience;
and (3) \emph{compatibility}, which implies that the proposed CAPTCHA generation
scheme is compatible with existing text- and image-based CAPTCHA deployment
and applications.

With the above goals in mind, we study the method to inject
human-tolerable, preprocessing-resilient (i.e., cannot be removed by CAPTCHA attacks)
perturbations to traditional CAPTCHAs.
Specifically, we design and implement a novel system \emph{aCAPTCHA}
to generate and evaluate text- and image-based
adversarial CAPTCHAs.

\vspace{2mm}
Our main contributions can be summarized as follows.

(1) Following our design principle, we propose a framework
for generating adversarial CAPTCHAs on top of existing adversarial
example (image) generation techniques. Specifically, we propose
four text-based and four image-based adversarial CAPTCHA generation
methods. Then, we design and implement a comprehensive
adversarial CAPTCHA generation and evaluation system, named \emph{aCAPTCHA},
which integrates 10 image preprocessing techniques,
9 CAPTCHA attacks, 4 baseline adversarial CAPTCHA generation methods,
and 8 new adversarial CAPTCHA generation methods.
aCAPTCHA can be used for the generation, security evaluation,
and usability evaluation of both text- and image-based
adversarial CAPTCHAs.

(2) To examine the performance of the adversarial CAPTCHAs
generated by aCAPTCHA, we conducted extensive security
and usability evaluations. The results demonstrate that
the generated adversarial CAPTCHAs can significantly improve the
security of normal CAPTCHAs while maintaining similar usability.

(3) We open source the aCAPTCHA system at \cite{acaptcha},
including the source code, trained models, datasets, and the interfaces
for usability evaluation. It is expected that aCAPTCHA
can facilitate the CAPTCHA security research and
can shed light on designing more secure and usable adversarial CAPTCHAs.


\section{Background} \label{backg}
In this section, we briefly introduce adversarial examples and the corresponding defense technologies.

\subsection{Adversarial Example}
Neural networks have achieved great performance on a wide range of application domains, especially, 
image recognition. However, recent work has discovered that the existing machine
learning models including neural networks are vulnerable to \emph{adversarial examples}. 
Specifically, suppose we have a classifier $F$ with model parameters $ \theta $. Let $x$ be an input
to the classifier with corresponding ground truth prediction $y$. An adversarial example 
$x'$ is an instance in the input space that is close to $x$ according to some distance metric 
$d(x,x')$, and causes classifier $F_\theta$ to produce an incorrect output.
Adversarial examples that affect one model often affect another model, even if the two models
have different architectures or were trained on different training sets, as long as both models
were trained to perform the same task\cite{papmcdarxiv16}. 

Prior work that considers adversarial examples under a number of threat models can be broadly 
classified into two categories: white-box attacks where the adversary has full knowledge of the model 
$F_\theta$ including the model architecture and parameters, and black-box attacks, where the adversary has 
no or little knowledge of the model $F_\theta$. The construction of an
adversarial example depends mainly on the gradient information of the target model. In the
white-box setting\cite{papmcdeurosp16,carwagsp17,gooshlcs14}, the gradient of the model is always visible to the attacker.
Thus, it is easy for an attacker to generate adversarial examples. In the black-box setting\cite{papmcdarxiv16,papmcdacccs17,ilyengICML2018}, attackers
cannot get gradient information directly. 
There are usually two ways to generate adversarial examples in this condition. The first one is to approximate the gradient 
information by query operations\cite{ilyengICML2018}, i.e., sending an image to the target model and getting the output distribution.
After many rounds of queries, attackers can approximate the target model's gradient and generate adversarial examples.
The second way is to take advantage of the transferability of adversarial examples\cite{papmcdarxiv16}. As we 
mentioned above, adversarial examples that affect one model can often affect another model.
An attacker could trains his own local model, generates adversarial examples against the local
model by white-box methods, and transfers them to a victim model which he has limited knowledge. 
In the paper, we rely on the second method refers to the black-box setting to generate adversarial CAPTCHAs against machine
learning based attacks. 

\subsection{Defense Methods}
Due to the security threats caused by adversarial examples,
improving the robustness of deep learning networks against adversarial perturbation has been an active 
field of research.
Various defensive techniques against adversarial examples
have been proposed. We roughly divide them into three categories.

\vspace{2mm}
(1) \emph{Adversarial Training\cite{gooshlcs14,trakuriclr2018}.}
The idea is simple and effective. One can retrain neural networks directly on adversarial examples
until the model learns to classify them correctly. 
This makes the network robust against the adversarial examples in the test 
set and improves the overall generalization capability of the network.  However,
it does not resolve the problem completely, as adversarial training can only be effective 
against specific adversarial example generation algorithms that are used in the retraining phase.
Moreover, adversarial training has been shown
to be difficult at a large scale, e.g., the ImageNet scale.

(2) \emph{Gradient Masking\cite{papmcdsp16,athcarICML2018}.}
This method tries to prevent an attacker from accessing the useful gradient information
of a model. As we mentioned, the construction of an
adversarial example depends mainly on the gradient information of the target model.
Without useful gradient information, the attackers are hard to perform an attack.
However, gradient masking is usually not effective
against black-box attacks, because an adversary could run his attack algorithm on an easy-to-attack model, 
and transfers these adversarial examples to the hard-to-attack model. 

(3) \emph{Input Transformation\cite{xuevandss18,bucroyiclr2018,guoraniclr18}.}
This kind of transformation method generally does not change the structure of a neural network.
The main idea is to preprocess or transform the input data, such as image cropping, rescaling and
bit-depth reduction, in order to remove adversarial
perturbation, and then feed the transformed image through an unmodified classifier.
This method is easy to 
circumvent by white-box attacks because attackers can
modify the attack algorithm in the mirror, e.g., considering similar operations during
adversarial examples generation. In the black-box attacks, it could provides good
protection. However, input transformation cannot eliminate adversarial
perturbation in the input data but only decreases the attack success rate.

In general, it is a fundamental problem that neural networks are vulnerable to adversarial perturbation. 
The existing defend methods are only to some extent mitigating the attack. Thus, dedicated in-depth research 
is expected in this area.

\section{System Overview} \label{sysoverview}

\begin{figure}[!tp]
	\centering
	\includegraphics[width=2.8in]{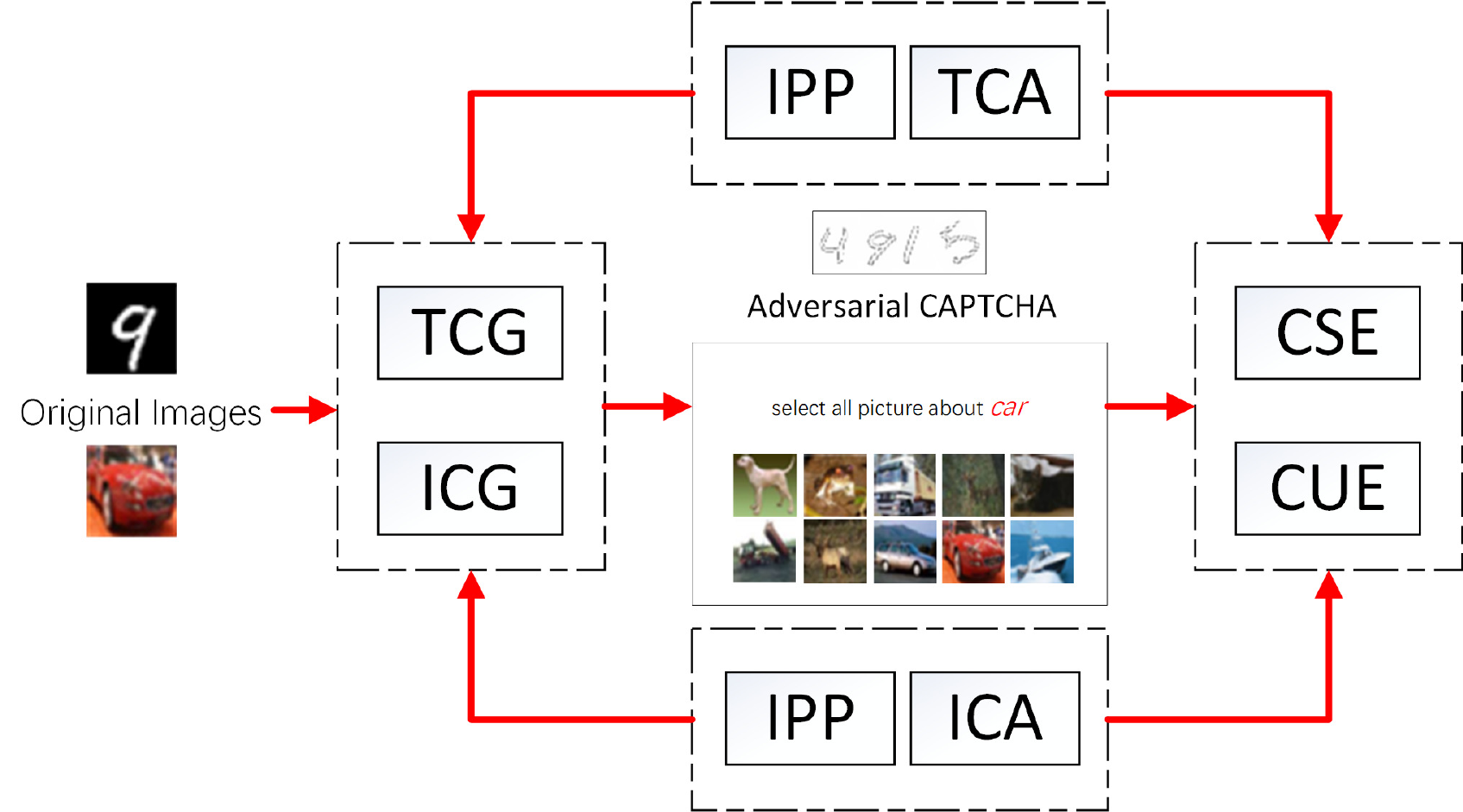}
	\caption{System overview of aCAPTCHA.} \label{f_overview}
\end{figure}

In this section, we present the system architecture of aCAPTCHA,
which is shown in Fig.\ref{f_overview}.
Basically, it consists of seven modules:

\vspace{2mm}
\textbf{Image Preprocessing (IPP) Module.}
In this module, we implement 10 widely used
standard image preprocessing techniques
for CAPTCHA security analysis, including 9 filters:
BLUR, DETAIL, EDGE ENHANCE,
SMOOTH, SMOOTH MORE, GaussianBlur, MinFilter,
MedianFilter, and ModeFilter,
and one standard image binarization method.
Basically, all the preprocessing techniques can be used
to remove the noise in an image.

\textbf{Text-based CAPTCHA Attack (TCA) Module.}
In this module, we implement 5 text-based CAPTCHA attacks,
including two traditional machine learning based attacks \emph{(SVM, KNN)} 
and three state-of-the-art DNN-based attacks 
\emph{(LeNet \cite{lecbotieee98}, MaxoutNet \cite{goowaricml13} and NetInNet \cite{lincheiclr13})}.
In aCAPTCHA, TCA has two main functions. First, it can provide
necessary model information for generating text-based adversarial CAPTCHAs,
i.e., for the following TCG module.
Second, it can also be employed to evaluate the resilience of
text-based CAPTCHAs against actual attacks.

\textbf{Image-based CAPTCHA Attack (ICA) Module.}
Similar to TCA, we implement 4 state-of-the-art image-based CAPTCHA attacks
in this module \emph{(NetInNet \cite{lincheiclr13}, VGG \cite{simzisarxiv14},
	GoogleNet \cite{szeliucvpr15} and ResNet \cite{hezhacvpr16})}. It is used to provide necessary model information
for generating image-based adversarial CAPTCHAs
and for evaluating the resilience of image-based CAPTCHAs
against actual attacks.

\textbf{Text-based Adversarial CAPTCHA Generation (TCG) Module.}
In this module,
we first implement 4 state-of-the-art adversarial example (image)
generation algorithms to serve as the baseline.
Then, we analyze the limitations of applying existing adversarial image
generation techniques to generate text-based adversarial CAPTCHAs.
Finally, according to our analysis, we propose 4 new
text-based adversarial CAPTCHA generation algorithms.

\textbf{Image-based Adversarial CAPTCHA Generation (ICG) Module.}
In this module, we first analyze the limitations of existing
adversarial image generation techniques for generating image-based adversarial CAPTCHAs.
Then, we implement 4 image-based adversarial CAPTCHA generation algorithms
by improving existing techniques.

\textbf{CAPTCHA Security Evaluation (CSE) Module.}
Leveraging TCA and ICA, this module is used to evaluate the
resilience and robustness of text- and image-based CAPTCHAs
against state-of-the-art attacks.

\textbf{CAPTCHA Usability Evaluation (CUE) Module.}
This module is mainly used for evaluating the usability of
text- and image-based CAPTCHAs.

aCAPTCHA takes a fully modular design, and is thus easily
extendable. We can freely add emerging attacks to TCA/ICA
and/or add new proposed adversarial CAPTCHA generation algorithms to
TCG/ICG.

\subsection{Datasets}

In the remainder of this paper, for the text-based evaluation scenario,
we employ
MNIST (Modified National Institute of Standards and Technology database)
\cite{mnist}.
MNIST is a large database of 70,000 handwritten digit images and is widely
used by the research community as a benchmark to evaluate text-based CAPTCHA's security
and usability \cite{gaoyanndss16}\cite{mnist}.

For the image-based evaluation scenario, we employ another image benchmark dataset
ImageNet ILSVRC-2012 (refers to the dataset used for 2012 ImageNet
Large Scale Visual Recognition Challenge)
\cite{imagenetwike}\cite{imagenet}.
The employed ImageNet ILSVRC-2012 contains 50,000 hand labeled
photographs from 1000 categories with 50 photographs from each category
\footnote{The used dataset here is a actually a subset of ImageNet ILSVRC-2012,
	which is sufficient for our purpose.}.

%
%

\section{Text-based Adversarial CAPTCHAs} \label{tcaptcha}

With the design goals in mind and following our design principle,
we show the design of TCG step by step below.

\subsection{Baselines}

In fact, CAPTCHAs can be viewed as a special case of images.
Then, following the design principle and goals, a straightforward
idea is to generate text-based adversarial CAPTCHAs using exiting
adversarial image generation techniques.
Therefore, we implement 4 baseline adversarial image
generation algorithms in TCG. Before delving to the details,
we define some useful notations.

\subsubsection{Notations}
We first present necessary notations
in the context of generating adversarial images.
To be consistent with existing research, we use the same notation system
as that in \cite{carwagsp17}.
We represent a neural network as a function $F(x) = y$,
where $x \in \mathbb{R}^{n\times n}$ is the input image \footnote{Note that,
	$x$ is not necessary to be a square image.
	The setting here is for simplicity.}
and $y \in \mathbb{R}^m$
is the corresponding output.
Define $F$ to be the full neural network including the softmax
function and let $Z(x) = z$ be the output of all the layers except the softmax.
According to $y$, $F$, which can be viewed as a classifier,
assigns $x$ a class label $C(x)$.
Let $C^*(x)$ be the correct label of $x$.

As in \cite{papmcdeurosp16}\cite{carwagsp17}, we use $L_p$ norms
to measure the similarity of $x, x' \in \mathbb{R}^{n\times n}$.
Then, $L_p = ||x - x'||_p = (\sum_{i = 1}^n \sum_{j = 1}^n |x - x'|^p)^{1/p}$.
According to the definition, $L_2$ distance measures the Euclidean distance
between $x$ and $x'$; $L_0$ distance measures the number of coordinates
$i$ s.t. $x_{i,j} \neq x_{i, j}'$; and $L_\infty$ distance measures
the maximum change to any of the coordinates, i.e.,
$||x - x'||_{\infty} = \max\{|x_{1, 1} - x_{1,1}'|, \cdots, |x_{n,n} - x_{n,n}'|\}$.

\subsubsection{Baseline Methods}
Recently, to generate adversarial examples (adversarial images in our context)
against neural networks, many attacks have been
proposed \cite{yuahearxiv17} \cite{akhmiaarxiv18}.
For our purpose, those attacks can serve as our adversarial CAPTCHA generation
methods. In TCG, we implement four state-of-the-art such attacks
as our baseline methods.

\vspace{2mm}
\textbf{JSMA.} In \cite{papmcdeurosp16}, Papernot et al. proposed the
Jacobian-based Saliency Map Attack (JSMA) to generate adversarial images.
JSMA is a greedy algorithm. Suppose $l$ is the target class of image $x$.
Then, to obtain $x'$ such that $x' \neq x$ and $C(x') = l$,
JSMA follows the following steps: (1) $x' = x$;
(2) based on the gradient $\nabla Z(x')_l$,
compute a \emph{saliency map} in which each value indicates the impact
of the corresponding pixel on the resulting classification;
(3) according to the saliency map, select the most important
pixel for modification to increase the likelihood of class $l$;
and (4) repeat the above two steps until $C(x') = l$
or more than a set threshold of pixels have been modified.

Note that, JSMA is also capable for generating untargeted adversarial
images. For that purpose, we only have to: (1) let $l = C(x)$
and change the goal as
to find $x'$ such that $x' \neq x$ and $C(x') \neq l$;
(2) select the pixel to mostly decrease the likelihood of class $l$
for modification.

\vspace{2mm}
\textbf{Carlini-Wagner Attacks.} Aiming at generating high quality
adversarial images, Carlini and Wagner in \cite{carwagsp17} introduced
three powerful attacks tailored to $L_2$, $L_0$, and $L_\infty$, respectively.
Basically, all those three attacks are optimization-based
and can be targeted or untargeted.
Taking the untargeted $L_2$ attack as an example, it can be formalized as
the optimization problem: minimize $||\delta|| + c \cdot F(x + \delta)$,
such that $x + \delta \in [0, 1]^{n}$,
i.e., for image $x$, the attack seeks for a perturbation $\delta$
that is small in length and can fool the classifier $F$ meanwhile.
In the formalization,
$c$ is a hyperparameter that balances the two parts in the objective function.
The constraint implies that the generated adversarial image should be valid.

\setlength{\tabcolsep}{3pt}
\begin{center}
	\begin{table*}[!tp]
		\fontsize{8pt}{9pt}\selectfont
		\caption{Performance of baseline algorithms vs LeNet.
			The original SAR of LeNet is $95.87\%$.} \label{tab_baseline}
		\centering
		\begin{tabular}{c  c  c  c  c  c  c  c  c}
			\hline
			\multirow{2}{*}{Filter} & \multicolumn{2}{c}{$JSMA$} & \multicolumn{2}{c}{$L_2$} & \multicolumn{2}{c}{$L_0$} & \multicolumn{2}{c}{$L_\infty$}\\
			& $-$ & $\mathbf{B}$ & $-$ & $\mathbf{B}$ & $-$ & $\mathbf{B}$ & $-$ & $\mathbf{B}$ \\
			\hline
			$-$
			& 0.00$\%$ & 13.93$\%$ & 0.00$\%$ & 73.51$\%$ & 0.00$\%$ & 1.38$\%$ & 0.00$\%$ & 83.30$\%$ \\
			BLUR
			& 5.15$\%$ & 8.27$\%$ & 4.22$\%$ & 20.84$\%$ & 6.25$\%$ & 19.27$\%$ & 6.25$\%$ & 22.52$\%$ \\
			DETAIL
			& 17.80$\%$ & 11.76$\%$ & 0.00$\%$ & 78.28$\%$ & 2.22$\%$ & 4.22$\%$ & 56.79$\%$ & 83.30$\%$ \\
			EDGE ENHANCE
			& 9.05$\%$ & 8.27$\%$ & 0.00$\%$ & 2.77$\%$ & 9.89$\%$ & 9.89$\%$ & 26.21$\%$ & 35.13$\%$ \\
			SMOOTH
			& 43.36$\%$ & 37.71$\%$ & 0.00$\%$ & 64.70$\%$ & 24.31$\%$ & 7.54$\%$ & 28.24$\%$ & 94.15$\%$ \\
			SMOOTH MORE
			& 37.71$\%$ & 40.46$\%$ & 0.00$\%$ & 37.71$\%$ & 20.84$\%$ & 10.79$\%$ & 19.27$\%$ & 88.58$\%$ \\
			GaussianBlur
			& 49.70$\%$ & 16.42$\%$ & 0.35$\%$ & 35.13$\%$ & 28.24$\%$ & 22.52$\%$ & 22.52$\%$ & 73.51$\%$ \\
			MinFilter
			& 0.15$\%$ & 1.38$\%$ & 0.05$\%$ & 0.11$\%$ & 0.02$\%$ & 0.07$\%$ & 0.06$\%$ & 0.15$\%$ \\
			MedianFilter
			& 24.31$\%$ & 68.99$\%$ & 0.05$\%$ & 35.13$\%$ & 17.80$\%$ & 12.81$\%$ & 12.81$\%$ & 68.99$\%$ \\
			ModeFilter
			& 20.84$\%$ & 30.40$\%$ & 0.00$\%$ & 22.52$\%$ & 30.40$\%$ & 32.69$\%$ & 0.05$\%$ & 40.46$\%$ \\
			\hline
		\end{tabular}
	\end{table*}
	\vspace{-6mm}
\end{center}

\subsection{Analysis of Baselines} \label{baseana}

As discussed before, intuitively, it seems like that existing adversarial image generation
algorithms, e.g., JSMA and Carlini-Wagner attacks, can be applied to
generate adversarial CAPTCHAs directly. Following this intuition,
we conduct a preliminary evaluation as follows:
($i$) Leveraging MNIST and standard CAPTCHA generation techniques \cite{captchanet},
randomly generate 10,000 CAPTCHAs of length 4, i.e., each CAPTCHA is composed
of 4 characters from MNIST; Denote these CAPTCHAs by set $\mathcal{C}$.
($ii$) Suppose LeNet from TCA is the employed CAPTCHA attack. Then, use LeNet
(trained using 50,000 CAPTCHAs for 20,000 rounds and
with batch size 50)
to attack the CAPTCHAs in $\mathcal{C}$. The \emph{Success Attack Rate} (SAR),
which is defined as the portion of successfully recognized CAPTCHAs in $\mathcal{C}$,
is $95.87\%$;
($iii$) In terms of LeNet, generate the adversarial versions of
the CAPTCHAs in $\mathcal{C}$ using JSMA, $L_2$, $L_0$, and $L_\infty$,
denoted by $\mathcal{C}_J$, $\mathcal{C}_2$, $\mathcal{C}_0$, and $\mathcal{C}_\infty$,
respectively.
($iv$) Use
LeNet  and possible preprocessing techniques from the IPP module
to attack $\mathcal{C}_J$, $\mathcal{C}_2$, $\mathcal{C}_0$, and $\mathcal{C}_\infty$.
The corresponding SARs are shown in Table \ref{tab_baseline},
where ``$-$" implies \emph{does not apply the corresponding preprocessing} and $\mathbf{B}$
denotes the \emph{image binarization} processing.

From Table \ref{tab_baseline}, we observe that
without applying image preprocessing, the adversarial CAPTCHAs
generated by all the baseline algorithms can significantly reduce
the SAR of LeNet, e.g., $L_2$ reduces the SAR of LeNet from
$95.87\%$ to $0\%$. This implies that the idea of applying adversarial CAPTCHAs
to defend against modern attacks is promising.

However, unfortunately, without talking the usability,
the security of these adversarial CAPTCHAs
can be significantly affected by image preprocessing either.
For instance, when attacking $\mathcal{C}_\infty$,
the SAR of LeNet is raised from $0\%$ to $28.24\%$ after applying
the SMOOTH filter and to $94.15\%$ after further applying image binarization,
which is similar to its performance on normal CAPTCHAs.
This implies that the perturbation in the adversarial CAPTCHAs
can be removed by image preprocessing, i.e., the perturbations
added by the baseline algorithms are not resilient/robust to image preprocessing.

We analyze the reasons from two aspects. 
From the perspective of breaking CAPTCHAs, text-based CAPTCHAs are monotonous
compared with the image-based CAPTCHAs. Character shape is only useful information
in text-based CAPTCHAs.
Other information, such as character colors and background pictures, is useless.
Thus, adversaries can employ multiple kinds of techniques,
e.g., filtering and image binarization, to remove noise and irrelevant information.
From the perturbation generation perspective,
theoretically, pre-processing such as filtering and binarization can be bypassed 
with minor modification of adversarial example generation algorithm, e.g., adding
another convolutional layer to the beginning of the neural network
with one output channel that performs similar filtering \cite{carwagarxiv2017}.
However, such modification will hugely increase the noise added in CAPTCHAs.
If we only consider filtering operation, the adversarial examples generated by
minor modification would not affect human recognition. While we consider both filtering 
and binarization, the adversarial examples generated by minor modification are unable
to recognize by human.
Therefore, existing adversarial image generation techniques cannot keep the balance 
between usability and security for text-based CAPTCHAs.

\subsection{Adversarial CAPTCHA Generation}

In the previous subsection, we analyzed the limitations of existing techniques
for generating adversarial CAPTCHAs.
Aiming at generating more robust and usable text-based adversarial CAPTCHAs,
we in this subsection proposed four new methods based on existing techniques.

Our design mainly follows two guidelines. First, according to our analysis,
the perturbations added in the space domain are frail to image preprocessing.
Therefore, we consider to add perturbations in the frequency domain.
This is because space domain perturbation can be considered as local change
of images while frequency domain perturbation is a kind of global change
to images, which is more difficult to remove, i.e., frequency domain perturbation
is intuitively more resilient to image preprocessing.
Certainly, when conducting frequency domain perturbation, we should be aware of
the possible impact on the usability.
Second, when generating adversarial CAPTCHAs, instead of trying to add human-imperceptible
perturbations, we focus on adding \emph{human-tolerable} perturbations.
This will give us more freedom to design more secure and fast
adversarial CAPTCHA generation methods. Specifically, based on JSMA, $L_2$, $L_0$, and $L_\infty$,
we propose 4 text-based adversarial CAPTCHA generation algorithms,
denoted by JSMA$^f$, $L_2^f$, $L_0^f$, and $L_\infty^f$, respectively.

\textbf{JSMA$^\textbf{\emph{f}}$.} We show the design of JSMA$^f$
in Algorithm \ref{a_jsmaf}. Basically,
JSMA$^f$ follows a similar procedure
as the untargeted JSMA. We remark the differences as follows.
First, in Steps 3-4, we transform a CAPTCHA to the frequency
domain by \emph{Fast Fourier Transform} (FFT) and then compute
a saliency map.
This enables us to elaborately inject perturbations to a CAPTCHA
in the frequency domain as expected.

Second, after transforming a CAPTCHA into the frequency domain,
its high frequency part usually corresponds to the margins of
characters and other non-vital information, while the low frequency part
usually corresponds to the fundamental shape information of characters.
Furthermore, as we indicated before, the changes made in the frequency domain
induce global changes to an image.
Therefore, to decrease possible impacts on the usability of a CAPTCHA,
we introduce a mask matrix $\varphi$ in Algorithm \ref{a_jsmaf},
which has the same size with $x$. $\varphi$ has values of 1
in the high frequency part while 0 in the low frequency part.
Then, as shown in Steps 5-6, we filter the pixels in the low frequency part
while only considering to change the pixels in the high frequency part.

Third, after selecting the candidate modified pixel,
instead of modifying one pixel each time as in JSMA,
we modify the candidate pixel and its neighbors as shown in Step 7.
This design is mainly based on the fact that close
pixels in the frequency domain exhibit the \emph{partial similarity} \cite{gonpearson08},
i.e., neighboring pixels in the frequency domain have very similar
property and features. Therefore, modifying the candidate pixel
and its neighbors would significantly accelerate the adversarial CAPTCHA
generation process while not harmfully affect its quality (recall that,
we are targeting to use user-tolerable instead of as little as possible perturbations).

Finally, we make an Inverse FFT (IFFT) for the CAPTCHA in the frequency domain
and transform it back to the space domain as shown in Step 8.

\vspace{2mm}

\textbf{$L_2^f$, $L_0^f$, and $L_\infty^f$.}
Basically, $L_2^f$, $L_0^f$, and $L_\infty^f$ follow the similar
procedures as that in $L_2$, $L_0$, and $L_\infty$ respectively, except that all
the designs are finished in the frequency domain.
The differences are the same as that between JSMA$^f$ and JSMA.
Therefore, we omit their algorithm descriptions here
while implementing them in TCG.

\begin{algorithm} [!tp] \label{a_jsmaf}
	\SetKwInOut{Input}{Input}
	\SetKwInOut{Output}{Output}
	\Input{$x$ original CAPTCHAs; $C^*(x)$ the label of $x$; $F$ a classifier; $\varphi$ mask.}
	\Output{$x'$ adversarial CAPTCHAs}
	\BlankLine
	
	$x' \leftarrow x$, $l \leftarrow C^*(x)$\;
	\While{$F(x') == l$}
	{
		$x'^f \leftarrow FFT(x')$\;
		compute a \emph{saliency map} $\mathcal{S}$ based on the gradient $\nabla Z(x'^f)_l$\;
		$\mathcal{S} \leftarrow \mathcal{S} \times \varphi$\;
		based on $\mathcal{S}$, select the pixel, denoted by $x'^f[i][j]$, that mostly decreases the likelihood of $l$\;
		modify $x'^f[i][j]$ and its neighbors to decrease the likelihood of $l$\;
		$x' \leftarrow IFFT(x'^f)$;
	}
	\caption{\textbf{JSMA$^f$}}
\end{algorithm}

\setlength{\tabcolsep}{3pt}
\begin{center}
	\begin{table*}[!tp]
		\fontsize{8pt}{9pt}\selectfont
		\caption{Performance of JSMA$^f$, $L_2^f$, $L_0^f$, and $L_\infty^f$ (no image preprocessing).}  \label{tab_tsec}
		\centering
		\begin{tabular}{c  c  c  c c c | c c c c | c c c c }
			\hline
			\multirow{3}{*}{Attack Model} & \multirow{3}{*}{ Normal} & \multicolumn{12}{c}{Text-based Adversarial CAPTCHA Generation} \\
			\cline{3-14}
			& & \multicolumn{4}{c|}{LeNet}  & \multicolumn{4}{c|}{MaxoutNet}& \multicolumn{4}{c}{NetInNet} \\
			\cline{3-14}
			\multicolumn{2}{c}{} & JSMA$^f$ & $L_2^f$  & $L_0^f$ & $L_\infty^f$ & JSMA$^f$ & $L_2^f$  & $L_0^f$ & $L_\infty^f$ & JSMA$^f$ & $L_2^f$  & $L_0^f$ & $L_\infty^f$ \\
			\hline
			\textbf{SVM}  & 87.51$\%$ & 0.00$\%$ & 0.00$\%$ & 0.00$\%$ & 0.00$\%$ & 0.00$\%$ & 0.00$\%$ & 0.00$\%$ & 0.00$\%$ & 0.00$\%$ & 0.00$\%$ & 0.00$\%$ & 0.00$\%$ \\
			\textbf{KNN} & 83.81$\%$ & 0.00$\%$ & 0.00$\%$ & 0.00$\%$ & 0.00$\%$ & 0.00$\%$ & 0.00$\%$ & 0.00$\%$ & 0.00$\%$ & 0.00$\%$ & 0.00$\%$ & 0.00$\%$ & 0.00$\%$ \\
			\textbf{LeNet} & 95.87$\%$ & 0.01$\%$ & 0.00$\%$ & 0.00$\%$ & 0.00$\%$ & 0.00$\%$ & 0.01$\%$ & 0.00$\%$ & 0.00$\%$ & 0.01$\%$ & 0.00$\%$ & 0.00$\%$ & 0.00$\%$ \\
			\textbf{MaxoutNet} & 95.29$\%$ & 0.00$\%$ & 0.00$\%$ & 0.00$\%$ & 0.00$\%$ & 0.00$\%$ & 0.00$\%$ & 0.00$\%$ & 0.00$\%$ & 0.00$\%$ & 0.00$\%$ & 0.00$\%$ & 0.00$\%$ \\
			\textbf{NetInNet}  & 96.45$\%$ & 0.00$\%$ & 0.00$\%$ & 0.00$\%$ & 0.00$\%$ & 0.00$\%$ & 0.00$\%$ & 0.00$\%$ & 0.00$\%$ & 0.00$\%$ & 0.00$\%$ & 0.00$\%$ & 0.00$\%$ \\
			\hline
		\end{tabular}
	\end{table*}
\end{center}

\setlength{\tabcolsep}{3pt}
\begin{center}
	\begin{table*}[!tp]
		\fontsize{8pt}{9pt}\selectfont
		\caption{Performance of JSMA$^f$, $L_2^f$, $L_0^f$, and $L_\infty^f$ (Filter + $\mathbf{B}$).} \label{tab_tsecfb}
		\centering
		\begin{tabular}{c c  c  c c c | c c c c | c c c c }
			\hline
			\multirow{3}{*}{Attack Model} & \multirow{3}{*}{ Filter + $\mathbf{B}$ } & \multicolumn{12}{c}{Text-based Adversarial CAPTCHA Generation} \\
			\cline{3-14}
			& & \multicolumn{4}{c|}{LeNet}  & \multicolumn{4}{c|}{MaxoutNet}& \multicolumn{4}{c}{NetInNet} \\
			\cline{3-14}
			\multicolumn{2}{c}{} & JSMA$^f$ & $L_2^f$  & $L_0^f$ & $L_\infty^f$ & JSMA$^f$ & $L_2^f$  & $L_0^f$ & $L_\infty^f$ & JSMA$^f$ & $L_2^f$  & $L_0^f$ & $L_\infty^f$ \\
			\hline
			\multirow{9}{*}{\rotatebox{90}{\textbf{SVM, KNN}}} &BLUR & 0.00$\%$ & 0.00$\%$ & 0.00$\%$ & 0.00$\%$ & 0.00$\%$ & 0.00$\%$ & 0.00$\%$ & 0.00$\%$ & 0.00$\%$ & 0.00$\%$ & 0.00$\%$ & 0.00$\%$ \\
			&DETAIL & 0.00$\%$ & 0.00$\%$ & 0.00$\%$ & 0.00$\%$ & 0.00$\%$ & 0.00$\%$ & 0.00$\%$ & 0.00$\%$ & 0.00$\%$ & 0.00$\%$ & 0.00$\%$ & 0.00$\%$ \\
			&EDGE ENHANCE & 0.00$\%$ & 0.00$\%$ & 0.00$\%$ & 0.00$\%$ & 0.00$\%$ & 0.00$\%$ & 0.00$\%$ & 0.00$\%$ & 0.00$\%$ & 0.00$\%$ & 0.00$\%$ & 0.00$\%$ \\
			&SMOOTH & 0.00$\%$ & 0.00$\%$ & 0.00$\%$ & 0.00$\%$ & 0.00$\%$ & 0.00$\%$ & 0.00$\%$ & 0.00$\%$ & 0.00$\%$ & 0.00$\%$ & 0.00$\%$ & 0.00$\%$ \\
			&SMOOTH MORE & 0.00$\%$ & 0.00$\%$ & 0.00$\%$ & 0.00$\%$ & 0.00$\%$ & 0.00$\%$ & 0.00$\%$ & 0.00$\%$ & 0.00$\%$ & 0.00$\%$ & 0.00$\%$ & 0.00$\%$ \\
			&GaussianBlur & 0.00$\%$ & 0.00$\%$ & 0.00$\%$ & 0.00$\%$ & 0.00$\%$ & 0.00$\%$ & 0.00$\%$ & 0.00$\%$ & 0.00$\%$ & 0.00$\%$ & 0.00$\%$ & 0.00$\%$ \\
			&MinFilter & 0.00$\%$ & 0.00$\%$ & 0.00$\%$ & 0.00$\%$ & 0.00$\%$ & 0.00$\%$ & 0.00$\%$ & 0.00$\%$ & 0.00$\%$ & 0.00$\%$ & 0.00$\%$ & 0.00$\%$ \\
			&MedianFilter & 0.00$\%$ & 0.00$\%$ & 0.00$\%$ & 0.00$\%$ & 0.00$\%$ & 0.00$\%$ & 0.00$\%$ & 0.00$\%$ & 0.00$\%$ & 0.00$\%$ & 0.00$\%$ & 0.00$\%$ \\
			&ModeFilter & 0.00$\%$ & 0.00$\%$ & 0.00$\%$ & 0.00$\%$ & 0.00$\%$ & 0.00$\%$ & 0.00$\%$ & 0.00$\%$ & 0.00$\%$ & 0.00$\%$ & 0.00$\%$ & 0.00$\%$ \\
			\hline
			\multirow{9}{*}{\rotatebox{90}{\textbf{LeNet}}} &BLUR & 0.32$\%$ & 0.29$\%$ & 0.26$\%$ & 0.24$\%$ & 0.30$\%$ & 0.49$\%$ & 0.30$\%$ & 0.28$\%$ & 0.38$\%$ & 0.25$\%$ & 0.35$\%$ & 0.29$\%$ \\
			&DETAIL & 3.77$\%$ & 0.48$\%$ & 1.98$\%$ & 1.84$\%$ & 2.71$\%$ & 4.01$\%$ & 2.86$\%$ & 2.77$\%$ & 3.32$\%$ & 2.24$\%$ & 3.47$\%$ & 2.95$\%$ \\
			&EDGE ENHANCE & 3.77$\%$ & 0.48$\%$ & 1.98$\%$ & 1.84$\%$ & 2.71$\%$ & 4.01$\%$ & 2.86$\%$ & 2.77$\%$ & 3.32$\%$ & 2.24$\%$ & 3.47$\%$ & 2.95$\%$ \\
			&SMOOTH & 11.66$\%$ & 3.50$\%$ & 6.19$\%$ & 6.56$\%$ & 8.49$\%$ & 10.89$\%$ & 7.20$\%$ & 7.47$\%$ & 10.70$\%$ & 8.12$\%$ & 8.65$\%$ & 7.97$\%$ \\
			&SMOOTH MORE & 8.89$\%$ & 2.71$\%$ & 5.10$\%$ & 4.81$\%$ & 6.94$\%$ & 9.13$\%$ & 5.68$\%$ & 5.85$\%$ & 8.49$\%$ & 6.49$\%$ & 6.81$\%$ & 6.56$\%$ \\
			&GaussianBlur & 0.03$\%$ & 0.05$\%$ & 0.04$\%$ & 0.04$\%$ & 0.04$\%$ & 0.07$\%$ & 0.05$\%$ & 0.04$\%$ & 0.05$\%$ & 0.06$\%$ & 0.05$\%$ & 0.04$\%$ \\
			&MinFilter & 0.00$\%$ & 0.00$\%$ & 0.00$\%$ & 0.00$\%$ & 0.00$\%$ & 0.00$\%$ & 0.00$\%$ & 0.00$\%$ & 0.00$\%$ & 0.00$\%$ & 0.00$\%$ & 0.00$\%$ \\
			&MedianFilter & 0.01$\%$ & 0.00$\%$ & 0.01$\%$ & 0.01$\%$ & 0.01$\%$ & 0.02$\%$ & 0.01$\%$ & 0.02$\%$ & 0.02$\%$ & 0.02$\%$ & 0.01$\%$ & 0.01$\%$ \\
			&ModeFilter & 0.01$\%$ & 0.00$\%$ & 0.01$\%$ & 0.01$\%$ & 0.01$\%$ & 0.02$\%$ & 0.01$\%$ & 0.02$\%$ & 0.02$\%$ & 0.02$\%$ & 0.01$\%$ & 0.01$\%$ \\
			\hline
			\multirow{9}{*}{\rotatebox{90}{\textbf{MaxoutNet}}} &BLUR & 5.85$\%$ & 5.68$\%$ & 4.67$\%$ & 5.20$\%$ & 5.63$\%$ & 3.89$\%$ & 5.63$\%$ & 5.15$\%$ & 5.96$\%$ & 8.34$\%$ & 5.46$\%$ & 5.20$\%$ \\
			&DETAIL & 10.70$\%$ & 3.85$\%$ & 7.61$\%$ & 6.87$\%$ & 8.57$\%$ & 8.81$\%$ & 8.73$\%$ & 8.57$\%$ & 10.15$\%$ & 6.25$\%$ & 9.80$\%$ & 9.21$\%$ \\
			&EDGE ENHANCE & 10.70$\%$ & 3.85$\%$ & 7.61$\%$ & 6.87$\%$ & 8.57$\%$ & 8.81$\%$ & 8.73$\%$ & 8.57$\%$ & 10.15$\%$ & 6.25$\%$ & 9.80$\%$ & 9.21$\%$ \\
			&SMOOTH & 38.25$\%$ & 28.66$\%$ & 31.53$\%$ & 29.96$\%$ & 37.98$\%$ & 34.88$\%$ & 35.13$\%$ & 34.88$\%$ & 37.45$\%$ & 35.13$\%$ & 35.38$\%$ & 34.88$\%$ \\
			&SMOOTH MORE & 38.52$\%$ & 27.83$\%$ & 30.85$\%$ & 30.85$\%$ & 34.88$\%$ & 32.69$\%$ & 33.89$\%$ & 34.38$\%$ & 36.92$\%$ & 31.53$\%$ & 34.14$\%$ & 33.89$\%$ \\
			&GaussianBlur & 0.13$\%$ & 0.47$\%$ & 0.14$\%$ & 0.16$\%$ & 0.16$\%$ & 0.05$\%$ & 0.12$\%$ & 0.12$\%$ & 0.14$\%$ & 0.31$\%$ & 0.14$\%$ & 0.14$\%$ \\
			&MinFilter & 0.00$\%$ & 0.00$\%$ & 0.00$\%$ & 0.00$\%$ & 0.00$\%$ & 0.00$\%$ & 0.00$\%$ & 0.00$\%$ & 0.00$\%$ & 0.00$\%$ & 0.00$\%$ & 0.00$\%$ \\
			&MedianFilter & 0.03$\%$ & 0.00$\%$ & 0.01$\%$ & 0.01$\%$ & 0.02$\%$ & 0.03$\%$ & 0.02$\%$ & 0.02$\%$ & 0.02$\%$ & 0.04$\%$ & 0.02$\%$ & 0.02$\%$ \\
			&ModeFilter & 0.03$\%$ & 0.00$\%$ & 0.01$\%$ & 0.01$\%$ & 0.02$\%$ & 0.03$\%$ & 0.02$\%$ & 0.02$\%$ & 0.02$\%$ & 0.04$\%$ & 0.02$\%$ & 0.02$\%$ \\
			\hline
			\multirow{9}{*}{\rotatebox{90}{\textbf{NetInNet}}} &BLUR & 17.51$\%$ & 13.47$\%$ & 14.64$\%$ & 15.77$\%$ & 17.10$\%$ & 16.29$\%$ & 16.55$\%$ & 15.26$\%$ & 18.08$\%$ & 17.10$\%$ & 16.82$\%$ & 16.03$\%$ \\
			&DETAIL & 15.90$\%$ & 7.34$\%$ & 10.89$\%$ & 10.42$\%$ & 13.47$\%$ & 14.52$\%$ & 13.03$\%$ & 12.49$\%$ & 13.47$\%$ & 10.24$\%$ & 13.59$\%$ & 13.03$\%$ \\
			&EDGE ENHANCE & 15.90$\%$ & 7.34$\%$ & 10.89$\%$ & 10.42$\%$ & 13.47$\%$ & 14.52$\%$ & 13.03$\%$ & 12.49$\%$ & 13.47$\%$ & 10.24$\%$ & 13.59$\%$ & 13.03$\%$ \\
			&SMOOTH & 28.24$\%$ & 16.82$\%$ & 19.89$\%$ & 20.20$\%$ & 24.49$\%$ & 24.87$\%$ & 22.01$\%$ & 21.84$\%$ & 24.12$\%$ & 21.67$\%$ & 23.04$\%$ & 22.18$\%$ \\
			&SMOOTH MORE & 28.88$\%$ & 19.42$\%$ & 21.84$\%$ & 21.33$\%$ & 24.87$\%$ & 26.21$\%$ & 23.40$\%$ & 22.35$\%$ & 24.87$\%$ & 22.18$\%$ & 23.58$\%$ & 22.52$\%$ \\
			&GaussianBlur & 0.48$\%$ & 0.28$\%$ & 0.27$\%$ & 0.22$\%$ & 0.44$\%$ & 0.43$\%$ & 0.37$\%$ & 0.35$\%$ & 0.45$\%$ & 0.64$\%$ & 0.39$\%$ & 0.37$\%$ \\
			&MinFilter & 0.00$\%$ & 0.00$\%$ & 0.00$\%$ & 0.00$\%$ & 0.00$\%$ & 0.00$\%$ & 0.00$\%$ & 0.00$\%$ & 0.00$\%$ & 0.00$\%$ & 0.00$\%$ & 0.00$\%$ \\
			&MedianFilter & 0.06$\%$ & 0.03$\%$ & 0.03$\%$ & 0.03$\%$ & 0.06$\%$ & 0.07$\%$ & 0.05$\%$ & 0.05$\%$ & 0.06$\%$ & 0.08$\%$ & 0.05$\%$ & 0.05$\%$ \\
			&ModeFilter & 0.06$\%$ & 0.03$\%$ & 0.03$\%$ & 0.03$\%$ & 0.06$\%$ & 0.07$\%$ & 0.05$\%$ & 0.05$\%$ & 0.06$\%$ & 0.08$\%$ & 0.05$\%$ & 0.05$\%$ \\
			\hline
		\end{tabular}
	\end{table*}
\end{center}

\subsection{Evaluation} \label{texteval}

Now, we evaluate the security performance of JSMA$^f$, $L_2^f$, $L_0^f$, and $L_\infty^f$
and leave their usability evaluation in Section \ref{usability}.
Generally, the evaluation procedure is the same as that in Section \ref{baseana}.
In all the evaluations of this subsection,
we employ MNIST to randomly generate CAPTCHAs of length 4.
For each attack in TCA, we use 50,000 normal CAPTCHAs for training.
Specifically, for the DNN based attacks LeNet, MaxOut, and NetInNet,
the batch size is 50 and each model is trained for 20,000 rounds.
For each scenario, we use 1000 CAPTCHAs for testing.
When generating an adversarial CAPTCHA, we set the
inner $8 \times 8$ area as the high frequency part while
the rest as the low frequency part for mask $\varphi$.
Each evaluation is repeated three times
and their average is reported as the final result.

First, we evaluate the performance of JSMA$^f$, $L_2^f$, $L_0^f$, and $L_\infty^f$
without  any image preprocessing. To conduct this group of evaluations,
we ($i$) leverage JSMA$^f$, $L_2^f$, $L_0^f$, and $L_\infty^f$
to generate adversarial CAPTCHAs in terms of LeNet, MaxoutNet, and NetInNet, respectively;
and ($ii$) leverage the attacks in the TCA module to attack these adversarial
CAPTCHAs, respectively. The results are shown in Table \ref{tab_tsec},
where \emph{Normal} indicates the SAR of each attack on the normal CAPTCHAs
(non-adversarial versions).

From Table \ref{tab_tsec}, we have the following observations.
(1) All the attacks in TCA are very powerful when attacking
normal CAPTCHAs. However, when they attack the adversarial CAPTCHAs
generated by JSMA$^f$, $L_2^f$, $L_0^f$, or $L_\infty^f$ ,
none of them can break any adversarial CAPTCHA.
This result is as expected and further demonstrates the advantage
of applying adversarial CAPTCHAs to improve the security.
(2) The generated CAPTCHAs by JSMA$^f$, $L_2^f$, $L_0^f$, and $L_\infty^f$
have very good transferability, i.e., the adversarial CAPTCHAs
generated in terms of one neural network model are transferable
to another neural network or traditional machine learning models.
This demonstrates the good robustness of the adversarial CAPTCHAs
generated by JSMA$^f$, $L_2^f$, $L_0^f$, and $L_\infty^f$.

Now, we go further by fully considering both image filtering and
image binarization, Common operations in breaking text-based
CAPTCHAs. Full results are shown in Table \ref{tab_tsecfb},
from which we have the following observations.
(1) For SVM and KNN, they cannot break any CAPTCHAs generated
by JSMA$^f$, $L_2^f$, $L_0^f$, or $L_\infty^f$ even after image preprocessing.
This implies adversarial CAPTCHAs can achieve very good
security when against traditional machine learning model
based attacks.
(2) For the DNN based attacks LeNet, MaxoutNet, and NetInNet,
they become more powerful along with
image filtering and binarization
and can break adversarial CAPTCHAs to some extent in several scenarios.
Still, adversarial CAPTCHAs are obviously more secure than normal ones
when considering the SAR rates of these attacks.
Further, comparing the results in Table \ref{tab_tsecfb} 
with that in Table \ref{tab_baseline}, the adversarial CAPTCHAs
generated by JSMA$^f$, $L_2^f$, $L_0^f$, and $L_\infty^f$
are also much more secure than the ones generated by
JSMA, $L_2$, $L_0$, and $L_\infty$.
(3) Similar as the previous evaluations, the adversarial CAPTCHAs
maintain adequate transferability,
which implies adversarial CAPTCHAs have stable robustness.

Finally, we discuss why the frequency-based methods perform better than 
space-based methods for text-based CAPTCHAs. According to the CAPTCHAs we generated 
(as shown in Fig. \ref{visual_image}), after adding noise in the frequency domain, 
the shape and edge of the character changes, which cannot be recovered by filtering and 
binaryzation. Furthermore, as we protect the low-frequency part of an image,
the fundamental shape of the characters in CHAPTCHAs will not change. Thus, human
can still recognize them easily.

%

\setlength{\tabcolsep}{1pt}
\begin{center}
	\begin{table*}[!tp]
		\fontsize{8pt}{9pt}\selectfont
		\caption{Security of image-based adversarial CAPTCHAs.} \label{tab_icap}
		\centering
		\begin{tabular}{c c  c  c  c c | c c c c | c c c c | c c c c   }
			\hline
			\multirow{3}{*}{} & \multirow{3}{*}{ Normal } & \multicolumn{15}{c}{Image-based Adversarial CAPTCHA Generation} \\
			\cline{3-18}
			& & \multicolumn{4}{c|}{NetInNet}  & \multicolumn{4}{c|}{GoogleNet}& \multicolumn{4}{c|}{VGG} & \multicolumn{4}{c}{ResNet50}  \\
			\cline{3-18}
			& & $JSMA^i$ & ${L_2}^i$ & ${L_0}^i$ & ${L_\infty}^i$ & $JSMA^i$ & ${L_2}^i$ & ${L_0}^i$ & ${L_\infty}^i$  & $JSMA^i$ & ${L_2}^i$ & ${L_0}^i$ & ${L_\infty}^i$  & $JSMA^i$ & ${L_2}^i$ & ${L_0}^i$ & ${L_\infty}^i$ \\
			\hline
			\textbf{NetInNet} & 41.72$\%$ & 0.0$\%$ & 0.0$\%$ & 0.0$\%$ & 0.0$\%$ & 4.6$\%$ & 20.3$\%$ & 4.2$\%$ & 1.9$\%$ & 0.7$\%$ & 5.4$\%$ & 1.9$\%$ & 2.2$\%$ & 4.1$\%$ & 3.3$\%$ & 8.8$\%$ & 4.7$\%$ \\
			\textbf{GoogleNet}& 51.69$\%$ & 0.5$\%$ & 3.8$\%$ & 7.0$\%$ & 14.3$\%$ & 0.0$\%$ & 0.0$\%$ & 0.0$\%$ & 0.0$\%$ & 0.0$\%$ & 0.1$\%$ & 0.2$\%$ & 1.5$\%$ & 0.4$\%$ & 1.2$\%$ & 6.3$\%$ & 4.6$\%$ \\
			\textbf{VGG}& 57.20$\%$ & 0.5$\%$ & 4.2$\%$ & 11.5$\%$ & 13.5$\%$ & 0.8$\%$ & 19.7$\%$ & 6.4$\%$ & 4.2$\%$ & 0.0$\%$ & 0.0$\%$ & 0.0$\%$ & 0.0$\%$ & 0.5$\%$ & 1.1$\%$ & 13.0$\%$ & 6.7$\%$ \\
			\textbf{ResNet50}& 63.80$\%$ & 10.1$\%$ & 17.6$\%$ & 18.5$\%$ & 20.8$\%$ & 1.9$\%$ & 26.2$\%$ & 7.9$\%$ & 7.6$\%$ & 0.1$\%$ & 0.4$\%$ & 1.2$\%$ & 3.1$\%$ & 0.0$\%$ & 0.0$\%$ & 0.0$\%$ & 0.0$\%$ \\
			\hline
		\end{tabular}
	\end{table*}
\end{center}

\setlength{\tabcolsep}{1pt}
\begin{center}
	\begin{table*}[!htp]
		\fontsize{8pt}{9pt}\selectfont
		\caption{Security of image-based adversarial CAPTCHAs vs Filters.} \label{tab_icapf}
		\centering
		\begin{tabular}{c c  c  c  c c | c c c c | c c c c | c c c c   }
			\hline
			\multirow{3}{*}{} & \multirow{3}{*}{ Filter } & \multicolumn{15}{c}{Image-based Adversarial CAPTCHA Generation} \\
			\cline{3-18}
			& & \multicolumn{4}{c|}{NetInNet}  & \multicolumn{4}{c|}{GoogleNet}& \multicolumn{4}{c|}{VGG} & \multicolumn{4}{c}{ResNet50}  \\
			\cline{3-18}
			& & $JSMA^i$ & ${L_2}^i$ & ${L_0}^i$ & ${L_\infty}^i$ & $JSMA^i$ & ${L_2}^i$ & ${L_0}^i$ & ${L_\infty}^i$  & $JSMA^i$ & ${L_2}^i$ & ${L_0}^i$ & ${L_\infty}^i$  & $JSMA^i$ & ${L_2}^i$ & ${L_0}^i$ & ${L_\infty}^i$ \\
			\hline
			\multirow{9}{*}{\rotatebox{90}{\textbf{NetInNet}}} &BLUR & 0.0$\%$ & 0.0$\%$ & 0.0$\%$ & 0.6$\%$ & 4.1$\%$ & 9.7$\%$ & 3.2$\%$ & 2.5$\%$ & 1.7$\%$ & 5.5$\%$ & 1.6$\%$ & 2.0$\%$ & 4.4$\%$ & 2.7$\%$ & 5.3$\%$ & 3.7$\%$ \\
			&DETAIL & 0.0$\%$ & 0.0$\%$ & 0.0$\%$ & 0.1$\%$ & 1.0$\%$ & 16.7$\%$ & 2.2$\%$ & 0.8$\%$ & 0.4$\%$ & 5.7$\%$ & 0.9$\%$ & 1.3$\%$ & 1.2$\%$ & 1.3$\%$ & 4.1$\%$ & 2.8$\%$ \\
			&EDGE ENHANCE & 0.0$\%$ & 0.0$\%$ & 0.0$\%$ & 0.0$\%$ & 0.0$\%$ & 2.0$\%$ & 0.0$\%$ & 0.0$\%$ & 0.0$\%$ & 0.0$\%$ & 0.0$\%$ & 0.0$\%$ & 0.0$\%$ & 0.0$\%$ & 0.1$\%$ & 0.0$\%$ \\
			&SMOOTH & 0.0$\%$ & 0.0$\%$ & 0.0$\%$ & 0.1$\%$ & 5.5$\%$ & 19.2$\%$ & 5.7$\%$ & 3.9$\%$ & 1.1$\%$ & 7.9$\%$ & 2.2$\%$ & 2.7$\%$ & 7.0$\%$ & 5.5$\%$ & 9.7$\%$ & 8.5$\%$ \\
			&SMOOTH MORE & 0.0$\%$ & 0.0$\%$ & 0.0$\%$ & 0.1$\%$ & 6.0$\%$ & 19.1$\%$ & 6.2$\%$ & 4.9$\%$ & 1.5$\%$ & 8.2$\%$ & 3.0$\%$ & 3.2$\%$ & 7.3$\%$ & 5.7$\%$ & 9.4$\%$ & 7.6$\%$ \\
			&GaussianBlur & 0.0$\%$ & 0.0$\%$ & 0.0$\%$ & 0.1$\%$ & 5.5$\%$ & 14.4$\%$ & 5.9$\%$ & 4.3$\%$ & 1.2$\%$ & 6.7$\%$ & 3.2$\%$ & 3.4$\%$ & 7.3$\%$ & 4.3$\%$ & 8.8$\%$ & 7.9$\%$ \\
			&MinFilter & 0.0$\%$ & 0.0$\%$ & 0.0$\%$ & 0.3$\%$ & 5.1$\%$ & 12.3$\%$ & 3.0$\%$ & 5.5$\%$ & 2.1$\%$ & 3.1$\%$ & 1.2$\%$ & 4.4$\%$ & 5.8$\%$ & 1.7$\%$ & 5.5$\%$ & 10.0$\%$ \\
			&MedianFilter & 0.0$\%$ & 0.0$\%$ & 0.0$\%$ & 0.1$\%$ & 5.1$\%$ & 16.2$\%$ & 6.5$\%$ & 2.9$\%$ & 1.2$\%$ & 7.9$\%$ & 3.5$\%$ & 2.3$\%$ & 5.5$\%$ & 5.7$\%$ & 9.4$\%$ & 5.3$\%$ \\
			&ModeFilter & 0.0$\%$ & 0.0$\%$ & 0.0$\%$ & 0.1$\%$ & 4.6$\%$ & 20.3$\%$ & 4.2$\%$ & 1.9$\%$ & 0.7$\%$ & 5.4$\%$ & 1.9$\%$ & 2.2$\%$ & 4.1$\%$ & 3.3$\%$ & 8.8$\%$ & 4.7$\%$ \\
			\hline
			\multirow{9}{*}{\rotatebox{90}{\textbf{GoogleNet}}} &BLUR & 0.8$\%$ & 5.7$\%$ & 5.9$\%$ & 8.2$\%$ & 0.0$\%$ & 3.9$\%$ & 1.3$\%$ & 1.3$\%$ & 0.9$\%$ & 4.3$\%$ & 2.8$\%$ & 2.9$\%$ & 6.8$\%$ & 3.4$\%$ & 6.2$\%$ & 6.5$\%$ \\
			&DETAIL & 0.4$\%$ & 2.9$\%$ & 5.1$\%$ & 12.6$\%$ & 0.0$\%$ & 0.0$\%$ & 0.0$\%$ & 0.0$\%$ & 0.0$\%$ & 0.1$\%$ & 0.2$\%$ & 1.2$\%$ & 0.2$\%$ & 0.5$\%$ & 3.1$\%$ & 3.0$\%$ \\
			&EDGE ENHANCE & 0.0$\%$ & 0.5$\%$ & 0.5$\%$ & 1.6$\%$ & 0.0$\%$ & 0.0$\%$ & 0.0$\%$ & 0.0$\%$ & 0.0$\%$ & 0.0$\%$ & 0.1$\%$ & 0.1$\%$ & 0.1$\%$ & 0.0$\%$ & 0.2$\%$ & 0.2$\%$ \\
			&SMOOTH & 0.3$\%$ & 2.7$\%$ & 5.7$\%$ & 7.8$\%$ & 0.0$\%$ & 0.0$\%$ & 0.0$\%$ & 0.0$\%$ & 0.0$\%$ & 0.8$\%$ & 0.7$\%$ & 1.5$\%$ & 0.9$\%$ & 1.7$\%$ & 7.9$\%$ & 5.0$\%$ \\
			&SMOOTH MORE & 0.4$\%$ & 3.8$\%$ & 7.1$\%$ & 9.1$\%$ & 0.0$\%$ & 0.0$\%$ & 0.0$\%$ & 0.0$\%$ & 0.0$\%$ & 0.8$\%$ & 0.8$\%$ & 1.5$\%$ & 1.1$\%$ & 1.2$\%$ & 8.1$\%$ & 5.7$\%$ \\
			&GaussianBlur & 0.4$\%$ & 2.3$\%$ & 4.3$\%$ & 6.2$\%$ & 0.0$\%$ & 0.8$\%$ & 0.0$\%$ & 0.6$\%$ & 0.1$\%$ & 2.1$\%$ & 2.1$\%$ & 2.0$\%$ & 1.9$\%$ & 3.2$\%$ & 9.1$\%$ & 6.0$\%$ \\
			&MinFilter & 2.1$\%$ & 3.8$\%$ & 4.2$\%$ & 10.8$\%$ & 0.0$\%$ & 1.2$\%$ & 0.2$\%$ & 1.7$\%$ & 0.2$\%$ & 1.1$\%$ & 0.7$\%$ & 3.4$\%$ & 1.7$\%$ & 0.7$\%$ & 5.3$\%$ & 7.6$\%$ \\
			&MedianFilter & 0.3$\%$ & 1.9$\%$ & 5.1$\%$ & 5.3$\%$ & 0.0$\%$ & 0.3$\%$ & 0.0$\%$ & 0.2$\%$ & 0.1$\%$ & 1.8$\%$ & 1.6$\%$ & 1.0$\%$ & 1.7$\%$ & 3.7$\%$ & 7.3$\%$ & 2.6$\%$ \\
			&ModeFilter & 0.5$\%$ & 3.8$\%$ & 7.0$\%$ & 14.3$\%$ & 0.0$\%$ & 0.0$\%$ & 0.0$\%$ & 0.0$\%$ & 0.0$\%$ & 0.1$\%$ & 0.2$\%$ & 1.5$\%$ & 0.4$\%$ & 1.2$\%$ & 6.3$\%$ & 4.6$\%$ \\
			\hline
			\multirow{9}{*}{\rotatebox{90}{\textbf{VGG}}} &BLUR & 1.0$\%$ & 4.9$\%$ & 5.5$\%$ & 7.6$\%$ & 3.7$\%$ & 15.2$\%$ & 8.5$\%$ & 5.3$\%$ & 0.0$\%$ & 0.0$\%$ & 0.0$\%$ & 0.3$\%$ & 2.2$\%$ & 3.4$\%$ & 7.6$\%$ & 6.7$\%$ \\
			&DETAIL & 0.8$\%$ & 4.2$\%$ & 11.5$\%$ & 10.8$\%$ & 0.5$\%$ & 18.1$\%$ & 5.3$\%$ & 2.0$\%$ & 0.0$\%$ & 0.0$\%$ & 0.0$\%$ & 0.0$\%$ & 0.2$\%$ & 0.7$\%$ & 10.4$\%$ & 3.1$\%$ \\
			&EDGE ENHANCE & 0.0$\%$ & 2.1$\%$ & 2.7$\%$ & 2.1$\%$ & 0.0$\%$ & 3.4$\%$ & 0.3$\%$ & 0.1$\%$ & 0.0$\%$ & 0.0$\%$ & 0.0$\%$ & 0.0$\%$ & 0.0$\%$ & 0.0$\%$ & 0.9$\%$ & 0.2$\%$ \\
			&SMOOTH & 0.7$\%$ & 3.9$\%$ & 9.2$\%$ & 13.1$\%$ & 2.6$\%$ & 21.3$\%$ & 11.8$\%$ & 4.7$\%$ & 0.0$\%$ & 0.0$\%$ & 0.0$\%$ & 0.0$\%$ & 1.2$\%$ & 3.1$\%$ & 13.9$\%$ & 8.0$\%$ \\
			&SMOOTH MORE & 0.7$\%$ & 3.7$\%$ & 10.0$\%$ & 12.3$\%$ & 2.0$\%$ & 20.8$\%$ & 11.5$\%$ & 6.2$\%$ & 0.0$\%$ & 0.0$\%$ & 0.0$\%$ & 0.0$\%$ & 1.2$\%$ & 3.2$\%$ & 13.5$\%$ & 9.4$\%$ \\
			&GaussianBlur & 1.1$\%$ & 4.8$\%$ & 7.6$\%$ & 10.4$\%$ & 3.9$\%$ & 21.8$\%$ & 11.9$\%$ & 5.3$\%$ & 0.0$\%$ & 0.0$\%$ & 0.0$\%$ & 0.0$\%$ & 2.1$\%$ & 3.4$\%$ & 13.5$\%$ & 7.5$\%$ \\
			&MinFilter & 2.9$\%$ & 3.0$\%$ & 4.7$\%$ & 8.8$\%$ & 4.5$\%$ & 10.1$\%$ & 3.2$\%$ & 7.6$\%$ & 0.0$\%$ & 0.0$\%$ & 0.0$\%$ & 0.3$\%$ & 4.9$\%$ & 1.6$\%$ & 5.4$\%$ & 12.3$\%$ \\
			&MedianFilter & 0.5$\%$ & 3.2$\%$ & 8.2$\%$ & 9.7$\%$ & 2.3$\%$ & 19.7$\%$ & 8.5$\%$ & 3.2$\%$ & 0.0$\%$ & 0.0$\%$ & 0.0$\%$ & 0.0$\%$ & 2.1$\%$ & 4.0$\%$ & 15.2$\%$ & 6.0$\%$ \\
			&ModeFilter & 0.5$\%$ & 4.2$\%$ & 11.5$\%$ & 13.5$\%$ & 0.8$\%$ & 19.7$\%$ & 6.4$\%$ & 4.2$\%$ & 0.0$\%$ & 0.0$\%$ & 0.0$\%$ & 0.0$\%$ & 0.5$\%$ & 1.1$\%$ & 13.0$\%$ & 6.7$\%$ \\
			\hline
			\multirow{9}{*}{\rotatebox{90}{\textbf{ResNet50}}} &BLUR & 4.1$\%$ & 17.6$\%$ & 10.8$\%$ & 14.6$\%$ & 6.2$\%$ & 24.0$\%$ & 11.1$\%$ & 6.7$\%$ & 0.9$\%$ & 7.0$\%$ & 6.0$\%$ & 4.4$\%$ & 0.1$\%$ & 0.5$\%$ & 2.6$\%$ & 4.6$\%$ \\
			&DETAIL & 10.8$\%$ & 16.7$\%$ & 15.6$\%$ & 18.6$\%$ & 1.3$\%$ & 22.3$\%$ & 5.1$\%$ & 4.1$\%$ & 0.2$\%$ & 0.5$\%$ & 1.1$\%$ & 2.4$\%$ & 0.0$\%$ & 0.0$\%$ & 0.0$\%$ & 0.0$\%$ \\
			&EDGE ENHANCE & 1.3$\%$ & 5.8$\%$ & 5.1$\%$ & 4.9$\%$ & 0.1$\%$ & 3.8$\%$ & 0.3$\%$ & 0.2$\%$ & 0.1$\%$ & 0.1$\%$ & 0.5$\%$ & 0.6$\%$ & 0.0$\%$ & 0.0$\%$ & 0.0$\%$ & 0.0$\%$ \\
			&SMOOTH & 6.2$\%$ & 13.9$\%$ & 15.2$\%$ & 17.1$\%$ & 4.1$\%$ & 33.1$\%$ & 14.2$\%$ & 7.3$\%$ & 0.0$\%$ & 0.4$\%$ & 1.0$\%$ & 3.1$\%$ & 0.0$\%$ & 0.0$\%$ & 0.0$\%$ & 0.0$\%$ \\
			&SMOOTH MORE & 7.0$\%$ & 14.4$\%$ & 17.1$\%$ & 19.6$\%$ & 4.3$\%$ & 30.4$\%$ & 15.6$\%$ & 7.8$\%$ & 0.1$\%$ & 1.0$\%$ & 1.2$\%$ & 3.0$\%$ & 0.0$\%$ & 0.0$\%$ & 0.0$\%$ & 0.0$\%$ \\
			&GaussianBlur & 4.4$\%$ & 16.7$\%$ & 12.3$\%$ & 15.6$\%$ & 8.2$\%$ & 30.6$\%$ & 13.5$\%$ & 8.8$\%$ & 0.1$\%$ & 3.8$\%$ & 2.9$\%$ & 4.5$\%$ & 0.0$\%$ & 0.0$\%$ & 0.2$\%$ & 1.0$\%$ \\
			&MinFilter & 7.0$\%$ & 9.3$\%$ & 10.1$\%$ & 16.9$\%$ & 10.8$\%$ & 16.6$\%$ & 7.0$\%$ & 10.8$\%$ & 1.0$\%$ & 2.1$\%$ & 1.8$\%$ & 4.6$\%$ & 0.1$\%$ & 0.1$\%$ & 2.1$\%$ & 3.9$\%$ \\
			&MedianFilter & 3.5$\%$ & 9.0$\%$ & 11.8$\%$ & 13.5$\%$ & 7.6$\%$ & 26.8$\%$ & 13.9$\%$ & 4.4$\%$ & 0.2$\%$ & 2.0$\%$ & 2.6$\%$ & 3.2$\%$ & 0.0$\%$ & 0.0$\%$ & 0.2$\%$ & 1.1$\%$ \\
			&ModeFilter & 10.1$\%$ & 17.6$\%$ & 18.5$\%$ & 20.8$\%$ & 1.9$\%$ & 26.2$\%$ & 7.9$\%$ & 7.6$\%$ & 0.1$\%$ & 0.4$\%$ & 1.2$\%$ & 3.1$\%$ & 0.0$\%$ & 0.0$\%$ & 0.0$\%$ & 0.0$\%$ \\
			\hline
		\end{tabular}
	\end{table*}
\end{center}

\section{Image-based Adversarial CAPTCHAs} \label{icaptcha}


\subsection{ICG Design}

For image-based adversarial CAPTCHA generation, we actually follow the same
design principles as that for the text-based scenario.
Furthermore, similar to the situation that existing adversarial image
generation techniques are not suitable for generating
text-based adversarial CAPTCHAs, they are not suitable for image-based
adversarial CAPTCHAs either due to similar reasons.
Existing adversarial image generation
techniques are mainly targeting to attack neural network models
by adding as less as possible (human-imperceptible) perturbations
to an image. However, we are standing on the defensive side to generate
adversarial CAPTCHAs to improve the security.
This implies that we might inject as much as possible perturbations
to an image-based adversarial CAPTCHA as long as it is user-tolerable
(user recognizable).
In addition, the adversarial example generation speed may not be a concern
for existing techniques. Although it is not a main constraint for
CAPTCHA generation neither, since we can generate the CAPTCHAs offline,
we still expect to generate many CAPTCHAs in a fast way (since we may need
to update our CAPTCHAs periodically to improve the system security).
Therefore, we take efficiency as a consideration in adversarial CAPTCHA
generation.

Image-based CAPTCHAs are also different from text-based ones.
They carry much richer semantic information which enables researchers
to develop more processing techniques.
Therefore, we do not have to transform an image-based CAPTCHA
to the frequency domain. To some extent, it is relatively easier
to generate image-based adversarial CAPTCHAs than generating
text-based adversarial CAPTCHAs.
Here, similar to the text-based scenario, we implement four
image-based adversarial CAPTCHA generation methods based on JSMA, $L_2$, $L_0$, and $L_\infty$,
denoted by JSMA$^i$, $L_2^i$, $L_0^i$, and $L_\infty^i$, respectively.

\begin{algorithm} [!tp] \label{a_jsmai}	
		\SetKwInOut{Input}{Input}
		\SetKwInOut{Output}{Output}
		\Input{$x$ original CAPTCHAs; $C^*(x)$ the label of $x$; $F$ a classifier; $K$ noise level.}
		\Output{$x'$ adversarial CAPTCHAs}
		\BlankLine
	$x' \leftarrow x$, $l \leftarrow C^*(x)$\;
	\While{$F(x') == l$ or $K > 0$ }
	{
		compute a \emph{saliency map} $\mathcal{S}$ based on the gradient $\nabla Z(x')_l$\;
		based on $\mathcal{S}$, select the pixel, denoted by $x'[i][j]$, that mostly decreases the likelihood of $l$\;
		modify $x'[i][j]$ and its neighbors to decrease the likelihood of $l$\;
		$K --$\;
	}

	\caption{\textbf{JSMA$^i$}}
\end{algorithm}

\vspace{2mm}
\textbf{JSMA$^\textbf{\emph{i}}$.}
We show the design of JSMA$^i$ in Algorithm \ref{a_jsmai},
which basically follows the same procedure as JSMA.
Following our design principle, we make two changes.
First, we introduce an integer parameter $K$ to control
the least perturbation that should be made.
This implies that in our design, we try to inject as much as possible
perturbations as long as the CAPTCHA is user tolerable
(certainly, $K$ is an empirical value that can be decided based on
some preliminary usability testing).
Second, like to the text-based scenario, we modify multiple
pixels simultaneously to accelerate the generation process.

\vspace{2mm}
\textbf{$L_2^i$, $L_0^i$, and $L_\infty^i$.}
For the designs of $L_2^i$, $L_0^i$, and $L_\infty^i$,
their procedures are the same as $L_2$, $L_0$, and $L_\infty$
except that we choose a small step and less iterations
to accelerate the CAPTCHA generation process.
This also implies that our perturbation injection
scheme may not be optimal compared with the original
$L_2$, $L_0$, and $L_\infty$. As we explained before,
we are not targeting to add as less perturbation as possible
like the original algorithms. Towards another direction,
we try to inject more perturbations in a fast way
when the CAPTCHA is user-tolerable.

\subsection{Evaluation}

Now, we evaluate the security performance of JSMA$^i$, $L_2^i$, $L_0^i$, and $L_\infty^i$
while leaving their usability evaluation in the next section.
In the evaluation, we employ ImageNet ILSVRC-2012 to generate all the needed
CAPTCHAs. Meanwhile, we use the pretrained models (all trained using the data
in ImageNet ILSVRC-2012)
of the attacks in ICA to examine
the security performance of the generated adversarial CAPTCHAs,
i.e., using the attacks in ICA to recognize the generated CAPTCHAs.
These pretrained models have state-of-the-art performance
and are available at Caffe Model Zoo \cite{caffemodelzoo}.
For each evaluation scenario, we use 1000 CAPTCHAs for testing.
Each evaluation is repeated three times and their average is reported as the final result.

\setlength{\tabcolsep}{2pt}
\begin{center}
	\begin{table*}[!tp]
		\fontsize{8pt}{9pt}\selectfont
		\caption{Security of image-based adversarial CAPTCHAs vs Noise level.} \label{tab_icapn}
		\centering
		\begin{tabular}{c c  c  c  c c | c cc c | c c c c | c c c c    }
			\hline
			\multirow{3}{*}{} & \multirow{3}{*}{ Filter } & \multicolumn{16}{c}{Image-based Adversarial CAPTCHA Generation} \\
			\cline{3-18}
			& & \multicolumn{4}{c|}{NetInNet}  & \multicolumn{4}{c|}{GoogleNet} & \multicolumn{4}{c|}{VGG}  & \multicolumn{4}{c}{ResNet50}   \\
			\cline{3-18}
			& & 20 & 30 & 40 & 50  & 20 & 30 & 40 & 50  & 20 & 30 & 40 & 50 & 20 & 30 & 40 & 50  \\
			\hline
			\multirow{9}{*}{\rotatebox{90}{\textbf{NetInNet}}}
			&BLUR & 0.0 $\%$ & 0.0 $\%$ & 0.0 $\%$ & 0.0 $\%$ & 1.6 $\%$ & 1.2 $\%$ & 1.1 $\%$ & 0.9 $\%$ & 0.6 $\%$ & 0.5 $\%$ & 0.4 $\%$ & 0.3 $\%$ & 2.7 $\%$ & 3.0 $\%$ & 2.4 $\%$ & 1.8 $\%$ \\
			&DETAIL & 0.0 $\%$ & 0.0 $\%$ & 0.0 $\%$ & 0.0 $\%$ & 0.8 $\%$ & 0.3 $\%$ & 0.1 $\%$ & 0.1 $\%$ & 0.2 $\%$ & 0.1 $\%$ & 0.1 $\%$ & 0.1 $\%$ & 0.9 $\%$ & 0.5 $\%$ & 0.2 $\%$ & 0.1 $\%$ \\
			&EDGE ENHANCE & 0.0 $\%$ & 0.0 $\%$ & 0.0 $\%$ & 0.0 $\%$ & 0.0 $\%$ & 0.0 $\%$ & 0.0 $\%$ & 0.0 $\%$ & 0.0 $\%$ & 0.0 $\%$ & 0.0 $\%$ & 0.0 $\%$ & 0.0 $\%$ & 0.0 $\%$ & 0.0 $\%$ & 0.0 $\%$ \\
			&SMOOTH & 0.0 $\%$ & 0.0 $\%$ & 0.0 $\%$ & 0.0 $\%$ & 2.7 $\%$ & 1.8 $\%$ & 1.2 $\%$ & 1.1 $\%$ & 0.3 $\%$ & 0.2 $\%$ & 0.2 $\%$ & 0.1 $\%$ & 3.6 $\%$ & 2.4 $\%$ & 1.6 $\%$ & 1.4 $\%$ \\
			&SMOOTH MORE & 0.0 $\%$ & 0.0 $\%$ & 0.0 $\%$ & 0.0 $\%$ & 3.0 $\%$ & 2.2 $\%$ & 1.2 $\%$ & 1.2 $\%$ & 0.4 $\%$ & 0.3 $\%$ & 0.3 $\%$ & 0.2 $\%$ & 3.6 $\%$ & 2.7 $\%$ & 2.2 $\%$ & 1.8 $\%$ \\
			&GaussianBlur & 0.0 $\%$ & 0.0 $\%$ & 0.0 $\%$ & 0.0 $\%$ & 1.8 $\%$ & 1.6 $\%$ & 1.4 $\%$ & 0.9 $\%$ & 0.3 $\%$ & 0.3 $\%$ & 0.1 $\%$ & 0.1 $\%$ & 3.3 $\%$ & 3.0 $\%$ & 1.8 $\%$ & 1.8 $\%$ \\
			&MinFilter & 0.0 $\%$ & 0.0 $\%$ & 0.0 $\%$ & 0.0 $\%$ & 2.0 $\%$ & 1.1 $\%$ & 1.2 $\%$ & 1.2 $\%$ & 0.3 $\%$ & 0.6 $\%$ & 0.4 $\%$ & 0.3 $\%$ & 2.0 $\%$ & 2.2 $\%$ & 2.0 $\%$ & 1.4 $\%$ \\
			&MedianFilter & 0.0 $\%$ & 0.0 $\%$ & 0.0 $\%$ & 0.0 $\%$ & 2.2 $\%$ & 1.8 $\%$ & 1.4 $\%$ & 1.2 $\%$ & 0.3 $\%$ & 0.4 $\%$ & 0.2 $\%$ & 0.1 $\%$ & 2.4 $\%$ & 1.8 $\%$ & 1.2 $\%$ & 1.1 $\%$ \\
			&ModeFilter & 0.0 $\%$ & 0.0 $\%$ & 0.0 $\%$ & 0.0 $\%$ & 1.8 $\%$ & 0.9 $\%$ & 0.6 $\%$ & 0.7 $\%$ & 0.2 $\%$ & 0.1 $\%$ & 0.1 $\%$ & 0.1 $\%$ & 2.4 $\%$ & 0.9 $\%$ & 0.5 $\%$ & 0.5 $\%$ \\
			\hline
			\multirow{9}{*}{\rotatebox{90}{\textbf{GoogleNet}}}
			&BLUR & 0.2 $\%$ & 0.2 $\%$ & 0.1 $\%$ & 0.1 $\%$ & 0.0 $\%$ & 0.0 $\%$ & 0.0 $\%$ & 0.0 $\%$ & 0.4 $\%$ & 0.1 $\%$ & 0.1 $\%$ & 0.1 $\%$ & 1.8 $\%$ & 2.0 $\%$ & 1.6 $\%$ & 1.8 $\%$ \\
			&DETAIL & 0.2 $\%$ & 0.1 $\%$ & 0.0 $\%$ & 0.0 $\%$ & 0.0 $\%$ & 0.0 $\%$ & 0.0 $\%$ & 0.0 $\%$ & 0.0 $\%$ & 0.0 $\%$ & 0.0 $\%$ & 0.0 $\%$ & 0.2 $\%$ & 0.1 $\%$ & 0.0 $\%$ & 0.0 $\%$ \\
			&EDGE ENHANCE & 0.0 $\%$ & 0.0 $\%$ & 0.0 $\%$ & 0.0 $\%$ & 0.0 $\%$ & 0.0 $\%$ & 0.0 $\%$ & 0.0 $\%$ & 0.0 $\%$ & 0.0 $\%$ & 0.0 $\%$ & 0.0 $\%$ & 0.0 $\%$ & 0.0 $\%$ & 0.0 $\%$ & 0.0 $\%$ \\
			&SMOOTH & 0.1 $\%$ & 0.1 $\%$ & 0.1 $\%$ & 0.0 $\%$ & 0.0 $\%$ & 0.0 $\%$ & 0.0 $\%$ & 0.0 $\%$ & 0.0 $\%$ & 0.0 $\%$ & 0.0 $\%$ & 0.0 $\%$ & 0.5 $\%$ & 0.1 $\%$ & 0.1 $\%$ & 0.1 $\%$ \\
			&SMOOTH MORE & 0.1 $\%$ & 0.1 $\%$ & 0.1 $\%$ & 0.1 $\%$ & 0.0 $\%$ & 0.0 $\%$ & 0.0 $\%$ & 0.0 $\%$ & 0.0 $\%$ & 0.0 $\%$ & 0.0 $\%$ & 0.0 $\%$ & 0.5 $\%$ & 0.3 $\%$ & 0.2 $\%$ & 0.1 $\%$ \\
			&GaussianBlur & 0.0 $\%$ & 0.1 $\%$ & 0.0 $\%$ & 0.0 $\%$ & 0.0 $\%$ & 0.0 $\%$ & 0.0 $\%$ & 0.0 $\%$ & 0.0 $\%$ & 0.0 $\%$ & 0.0 $\%$ & 0.0 $\%$ & 0.9 $\%$ & 0.3 $\%$ & 0.2 $\%$ & 0.2 $\%$ \\
			&MinFilter & 0.4 $\%$ & 0.3 $\%$ & 0.2 $\%$ & 0.2 $\%$ & 0.0 $\%$ & 0.0 $\%$ & 0.0 $\%$ & 0.0 $\%$ & 0.1 $\%$ & 0.0 $\%$ & 0.0 $\%$ & 0.0 $\%$ & 1.8 $\%$ & 0.9 $\%$ & 0.4 $\%$ & 0.2 $\%$ \\
			&MedianFilter & 0.1 $\%$ & 0.1 $\%$ & 0.1 $\%$ & 0.0 $\%$ & 0.0 $\%$ & 0.0 $\%$ & 0.0 $\%$ & 0.0 $\%$ & 0.0 $\%$ & 0.0 $\%$ & 0.0 $\%$ & 0.0 $\%$ & 0.6 $\%$ & 0.3 $\%$ & 0.3 $\%$ & 0.2 $\%$ \\
			&ModeFilter & 0.2 $\%$ & 0.1 $\%$ & 0.1 $\%$ & 0.0 $\%$ & 0.0 $\%$ & 0.0 $\%$ & 0.0 $\%$ & 0.0 $\%$ & 0.0 $\%$ & 0.0 $\%$ & 0.0 $\%$ & 0.0 $\%$ & 0.2 $\%$ & 0.1 $\%$ & 0.1 $\%$ & 0.0 $\%$ \\
			\hline
			\multirow{9}{*}{\rotatebox{90}{\textbf{VGG16}}}
			&BLUR & 0.4 $\%$ & 0.2 $\%$ & 0.1 $\%$ & 0.1 $\%$ & 1.6 $\%$ & 1.1 $\%$ & 0.9 $\%$ & 0.7 $\%$ & 0.0 $\%$ & 0.0 $\%$ & 0.0 $\%$ & 0.0 $\%$ & 0.8 $\%$ & 0.6 $\%$ & 0.6 $\%$ & 0.5 $\%$ \\
			&DETAIL & 0.1 $\%$ & 0.1 $\%$ & 0.1 $\%$ & 0.1 $\%$ & 0.3 $\%$ & 0.2 $\%$ & 0.1 $\%$ & 0.1 $\%$ & 0.0 $\%$ & 0.0 $\%$ & 0.0 $\%$ & 0.0 $\%$ & 0.2 $\%$ & 0.1 $\%$ & 0.0 $\%$ & 0.0 $\%$ \\
			&EDGE ENHANCE & 0.0 $\%$ & 0.0 $\%$ & 0.0 $\%$ & 0.0 $\%$ & 0.0 $\%$ & 0.0 $\%$ & 0.0 $\%$ & 0.0 $\%$ & 0.0 $\%$ & 0.0 $\%$ & 0.0 $\%$ & 0.0 $\%$ & 0.0 $\%$ & 0.0 $\%$ & 0.0 $\%$ & 0.0 $\%$ \\
			&SMOOTH & 0.3 $\%$ & 0.1 $\%$ & 0.1 $\%$ & 0.1 $\%$ & 0.9 $\%$ & 0.9 $\%$ & 0.7 $\%$ & 0.6 $\%$ & 0.0 $\%$ & 0.0 $\%$ & 0.0 $\%$ & 0.0 $\%$ & 0.6 $\%$ & 0.3 $\%$ & 0.1 $\%$ & 0.1 $\%$ \\
			&SMOOTH MORE & 0.2 $\%$ & 0.1 $\%$ & 0.1 $\%$ & 0.0 $\%$ & 0.9 $\%$ & 0.9 $\%$ & 0.5 $\%$ & 0.5 $\%$ & 0.0 $\%$ & 0.0 $\%$ & 0.0 $\%$ & 0.0 $\%$ & 0.3 $\%$ & 0.2 $\%$ & 0.1 $\%$ & 0.1 $\%$ \\
			&GaussianBlur & 0.3 $\%$ & 0.1 $\%$ & 0.1 $\%$ & 0.1 $\%$ & 1.2 $\%$ & 1.4 $\%$ & 0.9 $\%$ & 0.9 $\%$ & 0.0 $\%$ & 0.0 $\%$ & 0.0 $\%$ & 0.0 $\%$ & 0.9 $\%$ & 0.5 $\%$ & 0.3 $\%$ & 0.2 $\%$ \\
			&MinFilter & 1.6 $\%$ & 1.1 $\%$ & 0.9 $\%$ & 0.7 $\%$ & 1.6 $\%$ & 1.4 $\%$ & 1.2 $\%$ & 1.1 $\%$ & 0.0 $\%$ & 0.0 $\%$ & 0.0 $\%$ & 0.0 $\%$ & 1.8 $\%$ & 0.9 $\%$ & 0.8 $\%$ & 0.5 $\%$ \\
			&MedianFilter & 0.3 $\%$ & 0.1 $\%$ & 0.0 $\%$ & 0.0 $\%$ & 1.1 $\%$ & 0.8 $\%$ & 0.8 $\%$ & 0.6 $\%$ & 0.0 $\%$ & 0.0 $\%$ & 0.0 $\%$ & 0.0 $\%$ & 0.9 $\%$ & 0.5 $\%$ & 0.3 $\%$ & 0.3 $\%$ \\
			&ModeFilter & 0.1 $\%$ & 0.1 $\%$ & 0.1 $\%$ & 0.0 $\%$ & 0.4 $\%$ & 0.3 $\%$ & 0.1 $\%$ & 0.1 $\%$ & 0.0 $\%$ & 0.0 $\%$ & 0.0 $\%$ & 0.0 $\%$ & 0.3 $\%$ & 0.1 $\%$ & 0.1 $\%$ & 0.0 $\%$ \\
			\hline
			\multirow{9}{*}{\rotatebox{90}{\textbf{ResNet50}}}
			&BLUR & 1.6 $\%$ & 1.4 $\%$ & 1.2 $\%$ & 1.1 $\%$ & 3.0 $\%$ & 2.4 $\%$ & 2.0 $\%$ & 1.8 $\%$ & 0.6 $\%$ & 0.2 $\%$ & 0.2 $\%$ & 0.2 $\%$ & 0.0 $\%$ & 0.0 $\%$ & 0.0 $\%$ & 0.0 $\%$ \\
			&DETAIL & 3.6 $\%$ & 3.6 $\%$ & 3.3 $\%$ & 3.0 $\%$ & 0.6 $\%$ & 0.5 $\%$ & 0.3 $\%$ & 0.3 $\%$ & 0.1 $\%$ & 0.0 $\%$ & 0.0 $\%$ & 0.0 $\%$ & 0.0 $\%$ & 0.0 $\%$ & 0.0 $\%$ & 0.0 $\%$ \\
			&EDGE ENHANCE & 0.2 $\%$ & 0.1 $\%$ & 0.1 $\%$ & 0.1 $\%$ & 0.0 $\%$ & 0.0 $\%$ & 0.0 $\%$ & 0.0 $\%$ & 0.0 $\%$ & 0.0 $\%$ & 0.0 $\%$ & 0.0 $\%$ & 0.0 $\%$ & 0.0 $\%$ & 0.0 $\%$ & 0.0 $\%$ \\
			&SMOOTH & 1.6 $\%$ & 1.6 $\%$ & 1.2 $\%$ & 1.2 $\%$ & 1.8 $\%$ & 1.4 $\%$ & 1.4 $\%$ & 0.9 $\%$ & 0.0 $\%$ & 0.0 $\%$ & 0.0 $\%$ & 0.0 $\%$ & 0.0 $\%$ & 0.0 $\%$ & 0.0 $\%$ & 0.0 $\%$ \\
			&SMOOTH MORE & 2.0 $\%$ & 1.2 $\%$ & 1.1 $\%$ & 1.1 $\%$ & 1.8 $\%$ & 1.8 $\%$ & 1.4 $\%$ & 1.2 $\%$ & 0.0 $\%$ & 0.0 $\%$ & 0.0 $\%$ & 0.0 $\%$ & 0.0 $\%$ & 0.0 $\%$ & 0.0 $\%$ & 0.0 $\%$ \\
			&GaussianBlur & 1.2 $\%$ & 0.7 $\%$ & 0.7 $\%$ & 0.8 $\%$ & 3.0 $\%$ & 3.0 $\%$ & 2.0 $\%$ & 1.8 $\%$ & 0.0 $\%$ & 0.0 $\%$ & 0.0 $\%$ & 0.0 $\%$ & 0.0 $\%$ & 0.0 $\%$ & 0.0 $\%$ & 0.0 $\%$ \\
			&MinFilter & 3.3 $\%$ & 2.2 $\%$ & 2.4 $\%$ & 2.0 $\%$ & 5.9 $\%$ & 4.3 $\%$ & 3.6 $\%$ & 3.0 $\%$ & 0.2 $\%$ & 0.3 $\%$ & 0.1 $\%$ & 0.1 $\%$ & 0.0 $\%$ & 0.0 $\%$ & 0.0 $\%$ & 0.0 $\%$ \\
			&MedianFilter & 0.9 $\%$ & 0.8 $\%$ & 0.7 $\%$ & 0.7 $\%$ & 2.2 $\%$ & 2.2 $\%$ & 2.2 $\%$ & 2.2 $\%$ & 0.1 $\%$ & 0.1 $\%$ & 0.0 $\%$ & 0.0 $\%$ & 0.0 $\%$ & 0.0 $\%$ & 0.0 $\%$ & 0.0 $\%$ \\
			&ModeFilter & 2.4 $\%$ & 2.2 $\%$ & 2.4 $\%$ & 2.4 $\%$ & 0.8 $\%$ & 0.7 $\%$ & 0.6 $\%$ & 0.3 $\%$ & 0.1 $\%$ & 0.0 $\%$ & 0.0 $\%$ & 0.0 $\%$ & 0.0 $\%$ & 0.0 $\%$ & 0.0 $\%$ & 0.0 $\%$ \\
			\hline
		\end{tabular}
	\end{table*}
\end{center}

\vspace{-6mm}
We first evaluate the security of the adversarial CAPTCHAs
generated by JSMA$^i$, $L_2^i$, $L_0^i$, and $L_\infty^i$
in the scenario of not considering any image preprocessing.
The results are shown in Table \ref{tab_icap}.
\emph{Normal} implies the SAR of each attack when against normal CAPTCHAs,
and in the rest of evaluation scenarios,
we first generate adversarial CAPTCHAs in terms of the neural network model
of an attack, e.g., VGG, and then using different attacks to attack
them. Further, the default setting is $K = 50$ for JSMA$^i$,
and $K = 100$ for $L_2^i$, $L_0^i$, and $L_\infty^i$
(note that, in the original $L_2$, $L_0$, and $L_\infty$,
there is also a parameter to control the noise level.
We denote it by $K$ for consistence in $L_2^i$, $L_0^i$, and $L_\infty^i$).

From Table \ref{tab_icap}, we have the following observations.
First, for image-based CAPTCHAs, adversarial learning techniques
can significantly improve their security.
This further confirms our design principle:
\emph{to set one's own spear against one's own shield}.
Second, the generated adversarial CAPTCHAs
demonstrate adequate transferability, i.e.,
the adversarial CAPTCHAs generated in terms of one neural network
model also exhibits good resilience to other attacks.
Thus, they are robust.

Under the same settings with Table \ref{tab_icap}, we examine the
security performance of JSMA$^i$, $L_2^i$, $L_0^i$, and $L_\infty^i$
against the attacks in ICA plus image preprocessing.
Note that, since all the CAPTCHAs are color images,
we do not consider image binarization here.
We show the results in Table \ref{tab_icapf},
Basically, same conclusions can be drawn from Table \ref{tab_icapf} 
as that from Table \ref{tab_icap}.
In addition, we can find that image filtering has little impact
on the security of the adversarial CAPTCHAs generated by
JSMA$^i$, $L_2^i$, $L_0^i$, or $L_\infty^i$,
i.e., they are very robust.

Now, we consider the impact of different perturbation (noise) levels
on the security of the generated adversarial CAPTCHAs.
Taking JSMA$^i$ as an example, we show partial results in Table \ref{tab_icapn},
from which we make the following observations.
First, in most of the scenarios, when adding more noise,
better security can be achieved, which is consistent with
our intuition. However, according to the results,
such security improvement is slight in most of the cases.
Second, as before, the generated adversarial CAPTCHAs
are resilient and robust to various attacks.

%

\section{Adaptive Security Analysis} \label{adaptive}
In Sections \ref{tcaptcha} and \ref{icaptcha}, we evaluate the security 
performance of aCAPTCHA when attackers have no idea whether possible 
defense has been implemented.
In this section, we analyze in depth the adaptive methods that could
be applied against aCAPTCHA.

\subsection{Statement}
In practical scenario, we assume the threat follows all of the following models.
\begin{quote}
	
	\noindent\textbf{Knowledge of Adversarial Example Generation and Defense:} 
	The attacker has full knowledge of adversarial example generation
	and defense schemes. They can get that information from the research community
	and other means.\\
	\noindent\textbf{No Knowledge of CAPTCHA Generation:} The attacker can realize 
	that the CAPTCHAs were updated by adding adversarial noise, while they do 
	not know the specific model and method used to generate the adversarial CAPTCHAs.\\
	\noindent\textbf{No Access to the Source Images:}  The attacker
	can only access to all generated adversarial CAPTCHAs but
	not to their source. They has
	no knowledge about the particular image used for
	generating the adversarial CAPTCHAs.
\end{quote}

From the aCAPTCHAs generation perspective, we do not know which model
the attacker uses. From the attacker perspective, it is also reasonable to assume that
they do not know the specific
method we use. In summary, it is black-box attack versus black-box defense.

\subsection{Adaptive Attack}
When attackers are aware of the existence of the possible defense, they 
will try other state-of-the-art methods against adversarial CAPTCHAs. 
As we discussed in Section \ref{backg}, there are three types of defensive
techniques against adversarial examples:
adversarial training, gradient masking and input transformation.
Attackers can adopt these techniques to improve their attacks.
We introduce one representative method for each type of defense respectively below.

\vspace{2mm}

\textbf{Ensemble Adversarial Training} \cite{trakuriclr2018}:
This method augments a model's training data with adversarial examples crafted 
on other static pre-trained models. As a result, minimizing the training loss 
implies increased the robustness to black-box attacks from some set of models.
In particular, the model trained by this method won the first round of the
NIPS 2017 competition on Defenses against Adversarial Attacks. We believe this 
method is one of the most powerful choice against adversarial CAPTCHAs.

\vspace{2mm}
\textbf{Defense Distillation} \cite{papmcdsp16}:
This method is a type of gradient masking based defense technique.
Defensive distillation modifies the softmax function
to include a temperature constant $T$:
\begin{align}
softmax(x,T)_i = \frac{e^{x_i/T}}{\sum_je^{x_i/T}}
\end{align}
First, training a teacher model on the training set, using softmax at temperature $T$. 
Then using the teacher model to label each instance
in the training set with soft labels (the output vector from the teacher model),
using softmax at temperature $T$.
Finally, training the distilled model on the soft
labels from the teacher model, again using softmax at temperature $T$.  
Distillation can potentially increase the accuracy
on the test set as well as the robustness against adversarial examples.

\vspace{2mm}
\textbf{Thermometer Encoding} \cite{bucroyiclr2018}:
Actually, image binaryzation and filtering are representative instances of
input transformation \cite{xuevandss18}. In section \ref{tcaptcha}, we have demonstrated that our
text-based adversarial CAPTCHAs are resistant to them. Thus, we consider a more
effective method here.
In contrast to prior work which viewed adversarial examples as blind spot
in neural networks, Goodfellow et al. \cite{goobularxiv13} argued that the
reason adversarial examples exist is that neural networks 
behave in a largely linear manner. The purpose of thermometer
encoding is to break this linearity.
Given an image $x$, for each pixel color $x_{(i,j,c)}$, the $l$-level
thermometer encoding $\tau(x_{(i,j,c)})$ is a $l$-dimensional vector:
\begin{align}
\tau(x_{(i,j,c)}) = 
\begin{cases}
1 & \text{if  $x_{(i,j,c)} > k/l$}, \\
0 & \text{otherwise}. 
\end{cases}
\end{align}
For example, for a 10-level thermometer encoding, we had 
$\tau(0.57) = 1111100000$. Then we use thermometer encoding  to
train a model.


\setlength{\tabcolsep}{3pt}
\begin{center}
	\begin{table}
		\fontsize{8pt}{9pt}\selectfont
		\caption{Performance of adversarial CAPTCHAs against adaptive attack.} \label{tab_adaptivesec}
		\centering
		\begin{tabular}{c  c | c c c c  }
			\hline
			\multirow{2}{*}{ Adaptive Methods} & \multirow{2}{*}{ Normal } & \multicolumn{4 }{c}{Adversarial} \\
			\cline{3-6}
			& & JSMA$^f$ & $L_2^f$  & $L_0^f$ & $L_\infty^f$ \\
			\hline
			$-$ & 95.87$\%$ & 0.00$\%$ & 0.00$\%$ & 0.00$\%$ & 0.00$\%$ \\
			EnAdv. Training & 96.95$\%$& 28.40$\%$ & 21.26$\%$ & 25.84$\%$ & 23.20$\%$ \\
			EnAdv. Training$^+$ & 97.12$\%$& 48.61$\%$ & 41.37$\%$ & 39.35$\%$ & 41.37$\%$ \\
			Distillation & 94.36$\%$ & 5.35$\%$ & 4.96$\%$ & 6.77$\%$ & 5.13$\%$\\
			Therm. Encoding & 92.39$\%$& 12.19$\%$ & 7.68$\%$ & 9.89$\%$ & 11.24$\%$\\
			\hline
		\end{tabular}
	\end{table}
\end{center}

\subsection{Evaluation}
%
%
Generally, the evaluation procedure is the same as that in Section \ref{texteval}.
In all the evaluations of this subsection,
we employ MNIST to randomly generate CAPTCHAs of length 4.
For each scenario, we use 1000 CAPTCHAs for testing.
When generating an adversarial CAPTCHA, we set the
inner $8 \times 8$ area as the low frequency part while
the rest as the high frequency part for mask $\varphi$.
Each evaluation is repeated three times
and their average is reported as the final result.

Specifically, we use MaxoutNet to generate adversarial CAPTCHAs.
For ensemble adversarial training, we use MaxoutNet, NetInNet and LeNet to
generate adversarial examples by JSMA$^f$, $L_2^f$, $L_0^f$ and $L_\infty^f$ 
respectively, and use these examples to train a LeNet model. In Table \ref{tab_adaptivesec},
\emph{EnAdv. Training} means we do not use adversarial examples crafted on MaxoutNet, while
\emph{EnAdv. Training$^+$} do.
For defense distillation, we set $T$ as 100 which is the strong defense setting.
For Thermometer Encoding, we set $l$ as 16 which is the same as the original paper. 
In addition, image binaryzation is used in all of the tests. 

The results are shown in Table \ref{tab_adaptivesec},
from which we make the following observations.
First, defense distillation which is based on gradient masking is not suitable to black-box defense.
The result is  consistent with our analysis that gradient masking is not an effective solution against
black-box attacks.
Second, thermometer encoding shows limited value to recognize adversarial examples. This may be due to the
large perturbation we injected.
Third, ensemble adversarial training largely improves the SAR, especially in the \emph{EnAdv. Training$^+$} setting. 
However, in practice, attackers are hard to know what methods and models used in adversarial CAPTCHAs
generation, which restricts the practical effect of ensemble adversarial training.
Overall, the generated adversarial CAPTCHAs are resilient to state-of-the-art defense methods.

\subsection{Discussion}
Now, we would like to discuss why the results in this paper are better than 
previous work (the attacks based on the transferability of adversarial examples did not
perform well). 
First, we stand on the defense side,
and follow the rule to inject as much perturbation as possible
when the adversarial CAPTCHAs remain human-tolerable. 
Large perturbation magnitudes usually cause stronger defense effect.
Second, the recognition of CAPTCHAs is carried out by multiple recognition 	
tasks simultaneously. When the success rate of a single recognition task 
decreases, the overall success rate will decrease exponentially. For example, 
when the successful recognition rate of a single character is $50\%$, 
the expected successful recognition rate of the text-based CAPTCHAs of length four is $6.25\%$, 
and the expected successful recognition rate of the text-based CAPTCHAs of length six is 
only $1.5\%$.

Then we consider why the improved attacks based on state-of-the-art techniques is limited.
We inject larger perturbation into CAPTCHAs,
and input transformation, such as image rescaling and bit-depth reduction can only
eliminate part of the perturbation. As a result, the remaining perturbation 
is still effective to downgrade the recognition model.
Further, in this paper, we generate adversarial CAPTCHAs against the local model 
trained by ourselves, instead of attacking the target model directly.
There is no widely accepted conclusion about the phenomenon that an adversarial example
generated by one model is often misclassified by other models.
The existing adversarial example defense strategies cannot perform well against transfer attacks,
e.g., gradient masking based methods. Adversarial training, especially ensemble 
adversarial training, is regarded as the most effective defense strategy against 
black-box attacks. However, 
it requires the attacker to guess the methods and collect enough source images that
are used in adversarial CAPTCHAs generation, which implies a large potential cost.
Overall, existing adversarial example defense techniques are difficult, if not impossible, 
to break our adversarial CAPTCHAs.

In this section, we do not conduct further evaluation for image-based adversarial CAPTCHAs.
This is due to that training models on ImageNet require lots of computation resources. Furthermore, 
we believe that image-based adversarial CAPTCHAs are more secure than text-based adversarial CAPTCHAs.
On the one hand, image-based CAPTCHAs contain rich and important information which plays a key role
in image classification. Thus, attackers cannot use radical image preprocessing, such as 
image binarization, and this increases the dimensionality of the space of adversarial examples.
On the other hand, many state-of-the-art adversarial example detection techniques fail to or are hard to deploy on large-scale
datasets, e.g., ImageNet. This enhances the security of image-based adversarial CAPTCHAs 
indirectly.


\begin{figure} [h]
	\centering
	\subfigure[Text-based Adversarial CAPTCHAs]{
		\includegraphics[width=2in]{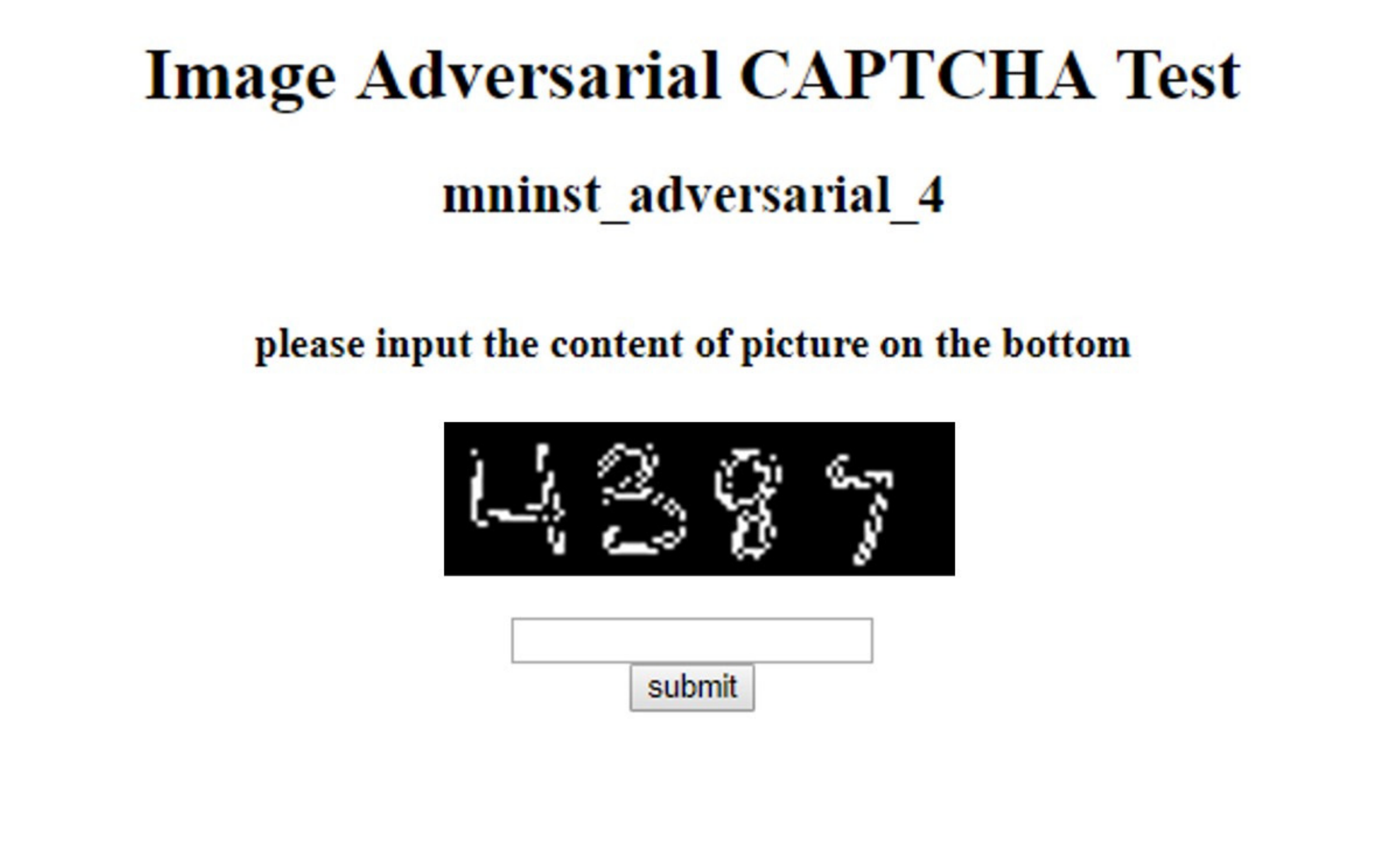}
	}
	\subfigure[Image-based Adversarial CAPTCHAs]{
		\includegraphics[width=3in]{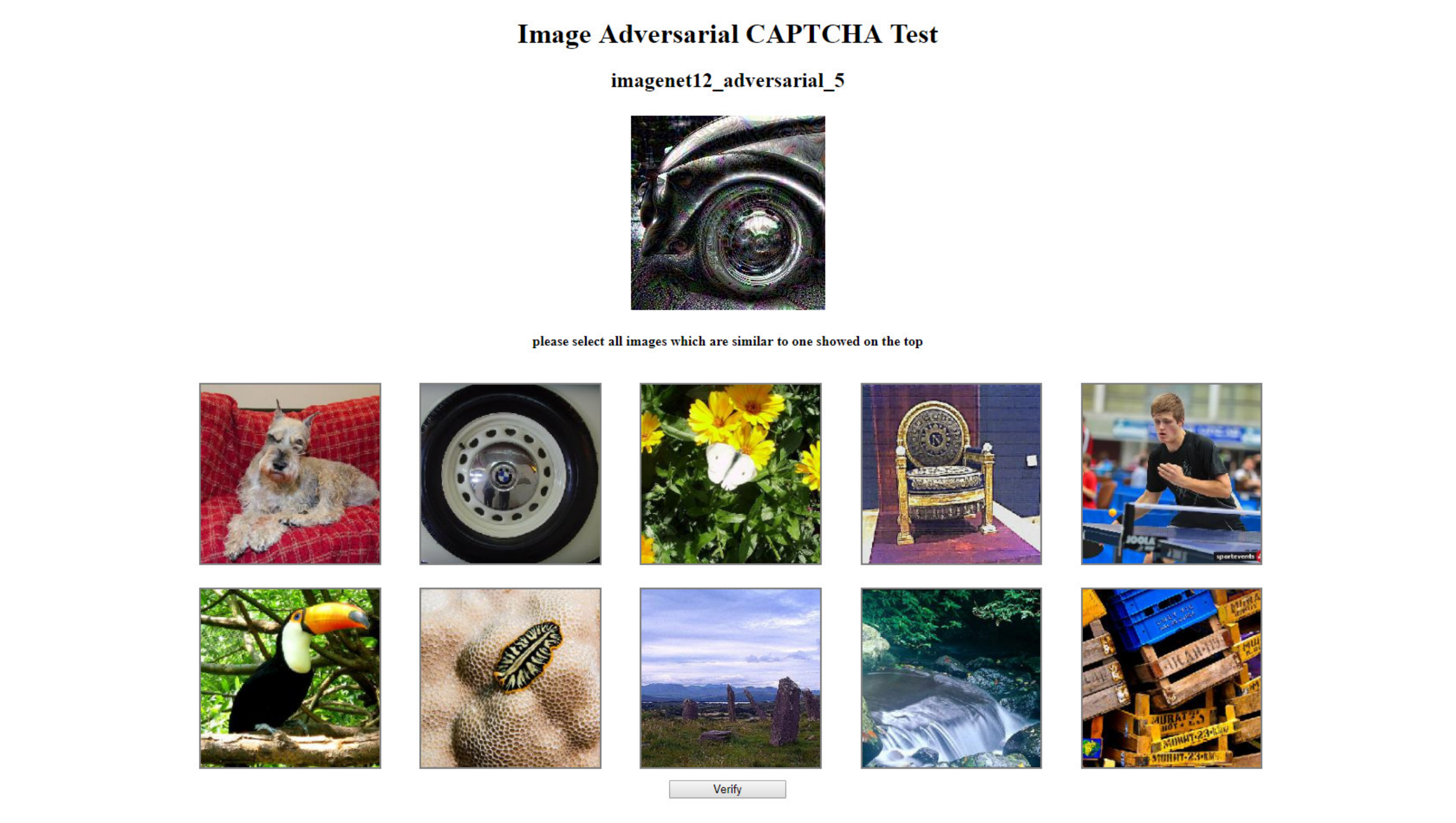}
	}
	\caption{Examples of aCAPTCHA. Text-based CAPTCHA is generated by JSMA$^f$ and
		image-based one is generated by JSMA$^i$ using K=50.} 
	\label{visual_image}
\end{figure} 


\section{Usability Evaluation} \label{usability}

We have examined the security performance of aCAPTCHA from multiple perspectives
in Sections \ref{tcaptcha}, \ref{icaptcha} and \ref{adaptive}, respectively.
In this section, we conduct experiments to evaluate the usability performance
of aCAPTCHA. As in the security evaluation, we employ MNIST and ImageNet ILSVRC-2012
to generate normal and adversarial CAPTCHAs for the text- and image-based
scenarios, respectively.

\subsection{Settings and Methodology}

To evaluate the usability of aCAPTCHA, we set the baseline as the usability
of normal text- and image-based CAPTCHAs.

\textbf{Methodology.}
To conduct our evaluation, we construct a real world website \cite{acaptcha},
on which the evaluation webpage is self-adapted to both PC and mobile clients, to
deploy normal and adversarial CAPTCHAs
and collect the evaluation data.
Then, we recruit volunteer users to do the evaluation.
For each user, she/he will be asked to finish the evaluation
in six steps.

Step 1: \emph{providing some general
	statistical information}, including gender, age range, and education level.

Step 2: \emph{finishing 10 tasks of recognizing randomly generated text-based normal CAPTCHAs},
including 5 CAPTCHAs of length 4
and 5 CAPTCHAs of length 6.

Step 3: \emph{finishing 10 tasks of recognizing randomly generated text-based adversarial CAPTCHAs},
including 5 CAPTCHAs of length 4 and 5 CAPTCHAs of length 6.
To simplify our evaluation, we here employ JSMA$^f$ to generate the adversarial CAPTCHAs.

Step 4: \emph{finishing 5 tasks of recognizing randomly generated image-based normal CAPTCHAs}.
For each recognition task, we first
randomly select two images belong to the same category from ILSVRC-2012,
and set one as the \emph{source image} and the other one as the \emph{target image}.
Then, we randomly select nine images from ILSVRC-2012
that are with different categories of the target image, and mix the target image
with the nine images to form a \emph{candidate set}.
Finally, given the source image, we ask a user to recognize
the target image from the candidate set.

Step 5: \emph{finishing 25 tasks of recognizing randomly generated
	image-based adversarial CAPTCHAs at five difficulty levels},
with each difficulty level has 5 tasks.
For each task in this step, its procedure is the same as the task in Step 4
except for the images used here are the adversarial versions.
For simplicity, we employ JSMA$^i$ (in terms of NetInNet) to generate the adversarial versions
for the source image and the images in the candidate set.
As shown in Section \ref{icaptcha}, we can control the noise level of JSMA$^i$
using $K$. Hence, in this step, we set up five difficulty levels
with $K = 10, 20, 30, 40, 50$, respectively. For each difficulty level,
each user is asked to do 5 tasks.

Step 6: \emph{providing some feedbacks of the evaluation}.
After finishing the previous five steps, we will show the user her/his
evaluation result, including how many tasks she/he failed,
which task she/he failed, etc. Then, we ask feedbacks from the users
by asking some questions, e.g., \emph{which CAPTCHA is more difficult to recognize?}

For each task in Steps 2 and 3, if all the characters in a CAPTCHA are correctly recognized,
we define that the task has been successfully finished.
For each task in Steps 4 and 5, if a user can correctly select the target image,
we define that the task has been
successfully finished.
After a user finished all the six steps, the results
will be transferred to the website server.
The visualization of adversarial CAPTCHAs used in test are shown in Figure \ref{visual_image}.

\textbf{Ethical Discussion.}
In our usability evaluation, human subjects are involved.
Therefore, we consulted with the IRB office for potential ethical issues.
Since we strictly limit ourselves to only collect necessary information
and no Personal Identifiable Information (PII) is collected,
our evaluation was approved by IRB.

\vspace{-12mm}

%

\setlength{\tabcolsep}{3pt}
\begin{center}
	\begin{table}[!tp]
		\fontsize{8pt}{9pt}\selectfont
		\caption{User statistics.} \label{tabusersta}
		\centering
		\begin{tabular}{cccccc}
			\hline
			\multirow{2}{*}{gender} & \multicolumn{2}{c}{female} & \multicolumn{3}{c}{male} \\
			\cline{2-6}
			& \multicolumn{2}{c}{43} & \multicolumn{3}{c}{82} \\
			\hline
			\multirow{2}{*}{age} & [16-20] & [21-30] & [31-40] & [41-50] & [51-60] \\
			\cline{2-6}
			& 76 & 40 & 1 & 3 & 5  \\
			\hline
			\multirow{2}{*}{education} & primary school & high school & B.S. & M.S. & Ph.D.\\
			\cline{2-6}
			& 1  & 17 & 85 & 12 & 10 \\
			\hline
		\end{tabular}
	\end{table}
\end{center}

\setlength{\tabcolsep}{3pt}
\begin{center}
	\begin{table*}[!tp]
		\fontsize{8pt}{9pt}\selectfont
		\caption{Usability of aCAPTCHA. } \label{tabusability}
		\centering
		\begin{tabular}{c  c  c | c  c | c | c  c  c  c  c}
			\hline
			& \multicolumn{4}{c |}{Text-based CAPTCHAs} & \multicolumn{6}{c}{Image-based CAPTCHAs} \\
			\cline{2-11}
			& \multicolumn{2}{c|}{Normal} & \multicolumn{2}{c|}{Adversarial} & \multirow{2}{*}{Normal} & \multicolumn{5}{c}{Adversarial} \\
			\cline{2-5}\cline{7-11}
			&  $\iota = 4$ &  $\iota = 6$ & $\iota = 4$ & $\iota = 6$ &  & $K = 10$ & $K = 20$ & $K = 30$ & $K = 40$ & $K = 50$ \\
			\hline
			Success rate & $92.8\%$ & $87.2\%$ & $88.0\%$ & $82.2\%$ & $80.0\%$ & $79.2\%$ & $81.6\%$ & $80.0\%$ & $80.8\%$ & $80.4\%$ \\
			Average time & 8.6s & 9.7s & 8.6s & 10.2s & 19.7s & 15.3s & 12.3s & 12.8s & 11.8s & 11.5s \\
			Median time & 7.1s & 7.8s & 6.2s & 8.4s & 16.0s & 10.9s & 9.4s & 9.4s & 8.8s & 9.4s \\
			\hline
		\end{tabular}
	\end{table*}
\end{center}

\subsection{Results and Analysis}

After moving the usability evaluation website online, we finally
recruit 125 volunteer users as shown in Table \ref{tabusersta}.
Specifically, the users include 43 females and 82 males,
and most of them have ages ranging from 16 to 30.
Furthermore, almost all the users' education levels
are high school or higher. Following the evaluation procedure,
all the 125 users successfully finished the evaluation
($\sim 90\%$ users finish the evaluation through smart phones).
We then collect all the results to our server.

Based on the collected data, we show the main results in Table \ref{tabusability},
where $\iota$ denotes the length of
a text-based CAPTCHA, $K$ indicates the noise (difficulty) level
of an image-based adversarial CAPTCHA, and
\emph{success rate}, \emph{average time},
and \emph{median time} measure the average successful probability,
the average time consumption, and the median time consumption
of all the users to finish the corresponding task, respectively.
From Table \ref{tabusability}, we have the following observations.

For text-based CAPTCHAs, although the adversarial versions
can significantly improve the security performance as shown in Section \ref{tcaptcha},
their success rate of recognition also maintains a high level, which is only slightly lower
than that of the normal versions. Meanwhile, it takes similar time
for users to recognize normal and adversarial CAPTCHAs.
These results suggest that text-based adversarial and normal CAPTCHAs
have similar usability.
In addition, given that long CAPTCHAs usually have better
security than short ones \cite{burmarccs11}, we also find that long text-based CAPTCHAs cost more
time for recognition and have a lower success rate than that of the short ones (consistent with our intuition).
This  implies that there is a tradeoff between security and usability.

\begin{figure}[!tp]
	\centering
	\subfigure[Success rate VS Gender]{
		\includegraphics[width=2.4in]{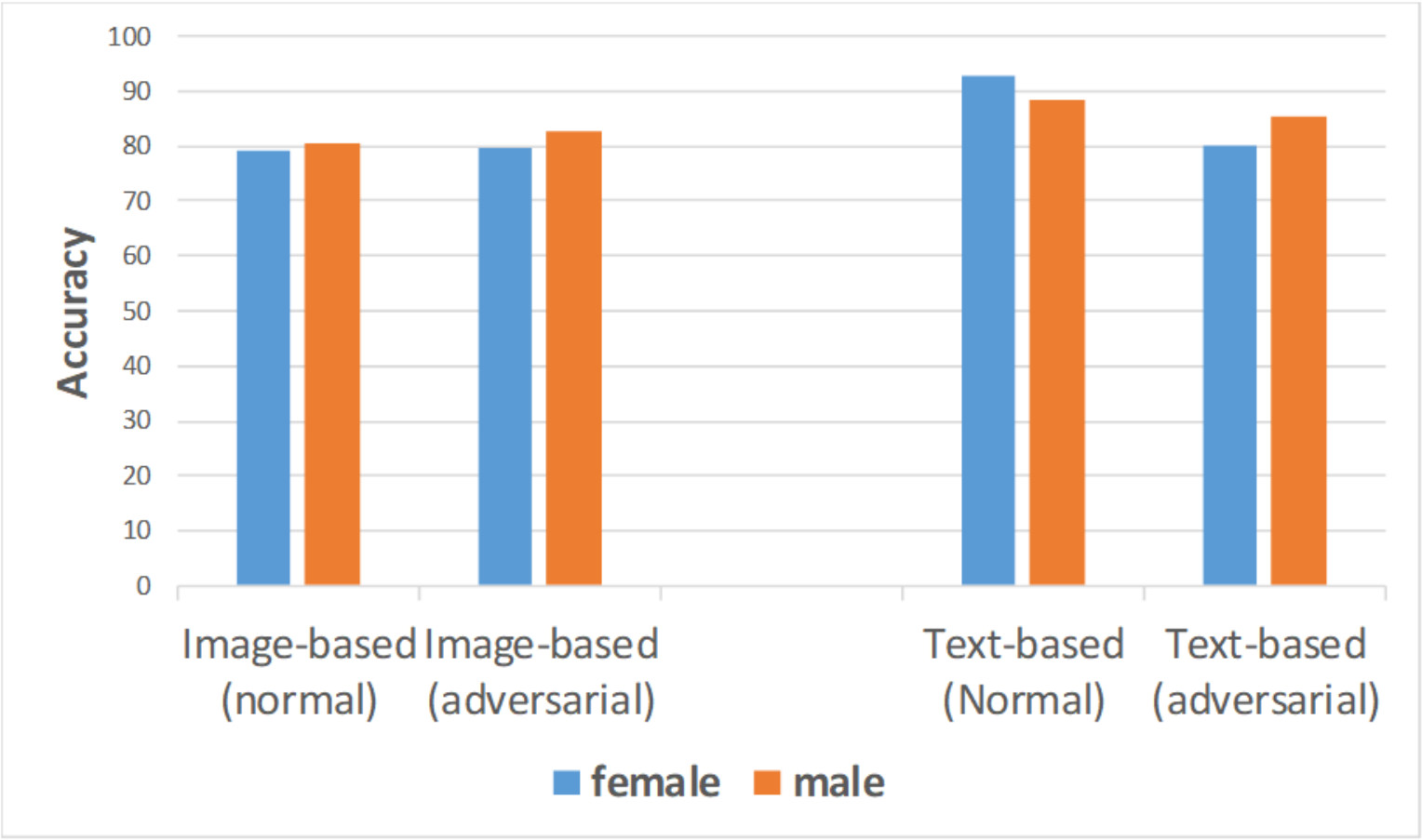}
	}
	\subfigure[Success rate VS Age]{
		\includegraphics[width=2.4in]{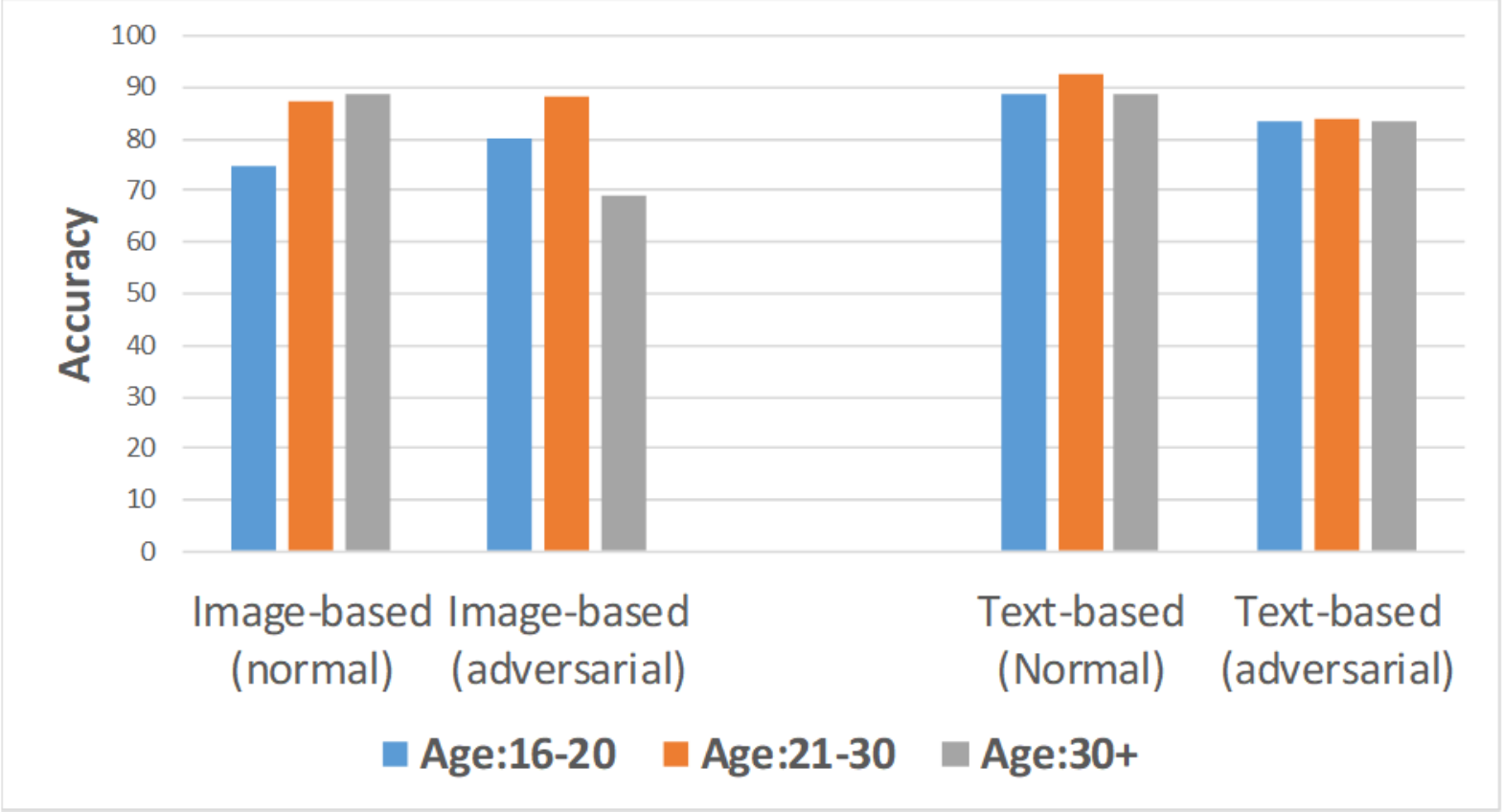}
	}
	\subfigure[Success rate VS Education]{
		\includegraphics[width=2.4in]{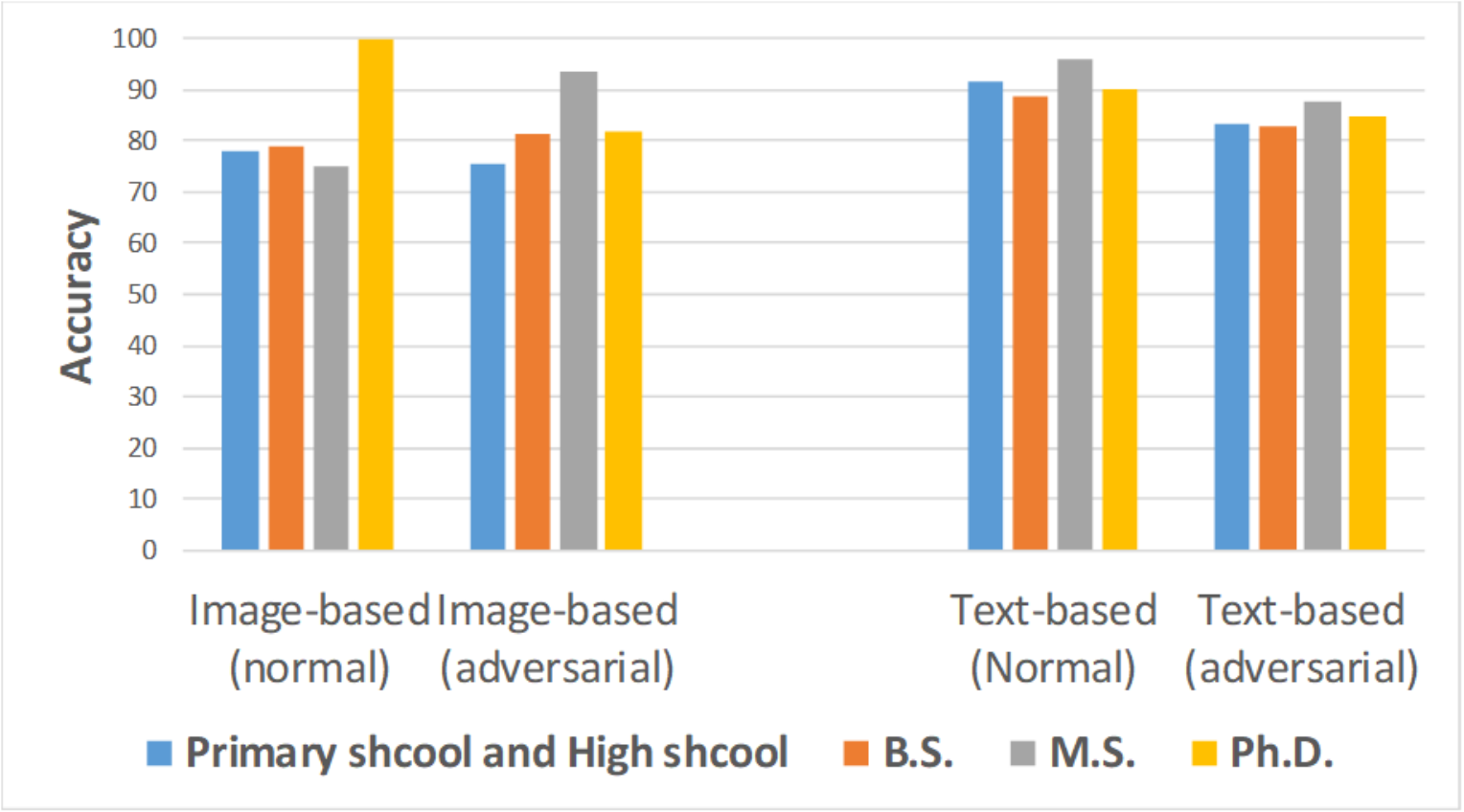}
	}
	\caption{Success rate VS Statistical category.} \label{f_catsuccess}
\end{figure}

For image-based CAPTCHAs, the advantage of adversarial versions
is more evident. Adversarial CAPTCHAs have similar or even better success rates
as the normal ones in all the cases. The success rates of adversarial CAPTCHAs
with different noise (difficulty) levels are also similar.
This suggests that image-based CAPTCHAs are more robust to adversarial perturbations.
Given the obvious security advantage shown
in Section \ref{icaptcha}, image-based adversarial CAPTCHAs is more promising compared
to normal ones. Another interesting observation is that adversarial CAPTCHAs
cost less time for recognition than the normal versions, which is a little bit
out of our expectation.
We conjecture the reasons as follows: ($i$) deliberately adversarial perturbation has little
impact on the quality of images with respect to human recognition;
and ($ii$) as the evaluation goes on, users become more and more familiar with the tasks.
Thus, they can finish the tasks faster.

Now, we give a close look at the success rate of different users
based on their statistical categories. The results are shown in Fig.\ref{f_catsuccess}.
From Fig.\ref{f_catsuccess}, we can see that, in most of the scenarios,
users from different statistical categories exhibit similar success rate
over both adversarial and normal CAPTCHAs.
This further demonstrates the generality of aCAPTCHA.


In summary, according to our evaluation, the CAPTCHAs generated by aCAPTCHA,
especially the image-based adversarial CAPTCHAs,
have similar usability as the normal versions.
Recall the security evaluation of aCAPTCHA in Sections \ref{tcaptcha} and \ref{icaptcha},
they together demonstrate that \emph{aCAPTCHA is promising
	in addressing the dilemma of existing text- and image-based CAPTCHAs}.


\begin{figure}[!tp]
	\centering
	\subfigure[Text-based CAPTCHAs]{
		\includegraphics[width=2.5in]{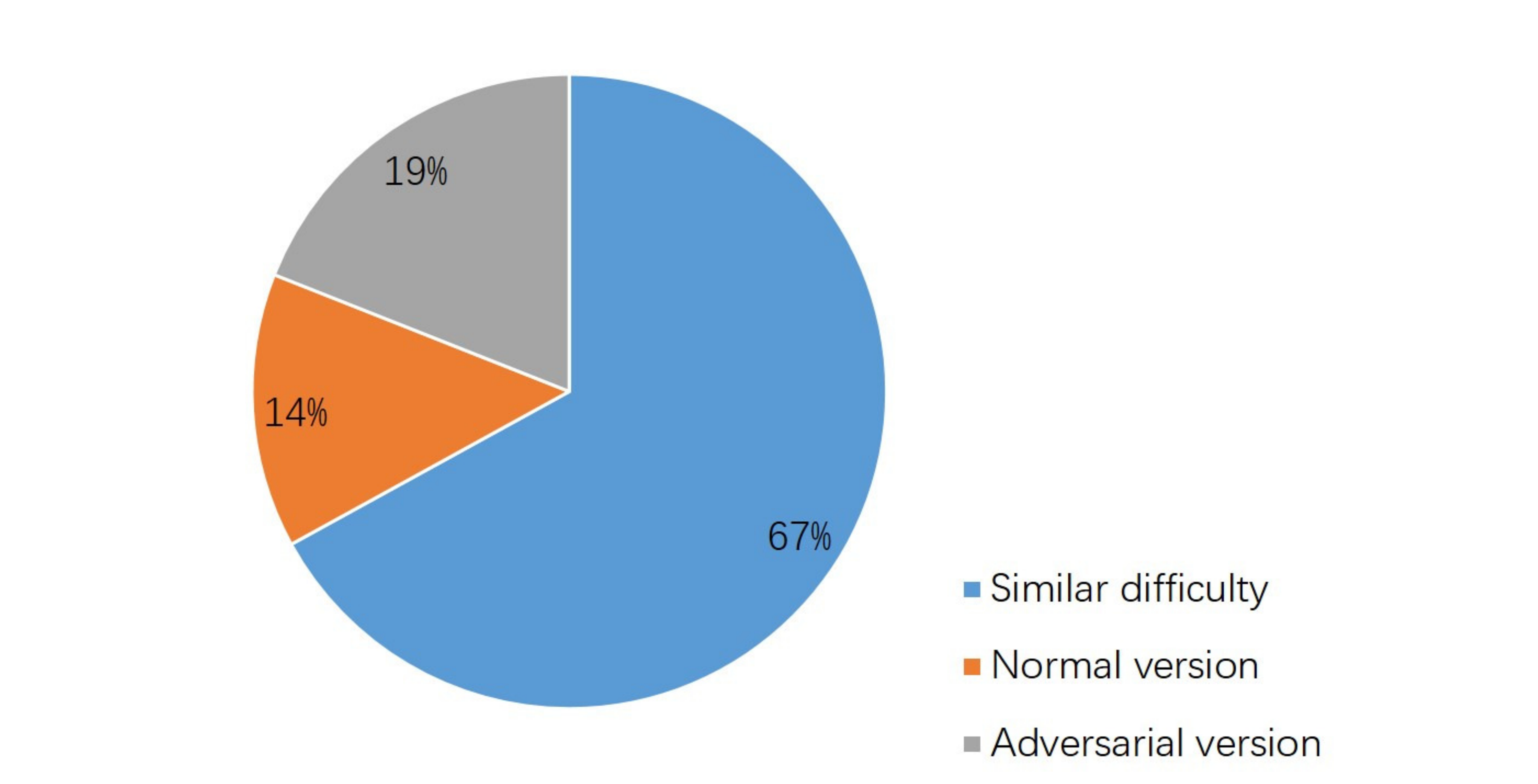}
	}
	\subfigure[Image-based CAPTCHAs]{
		\includegraphics[width=2.5in]{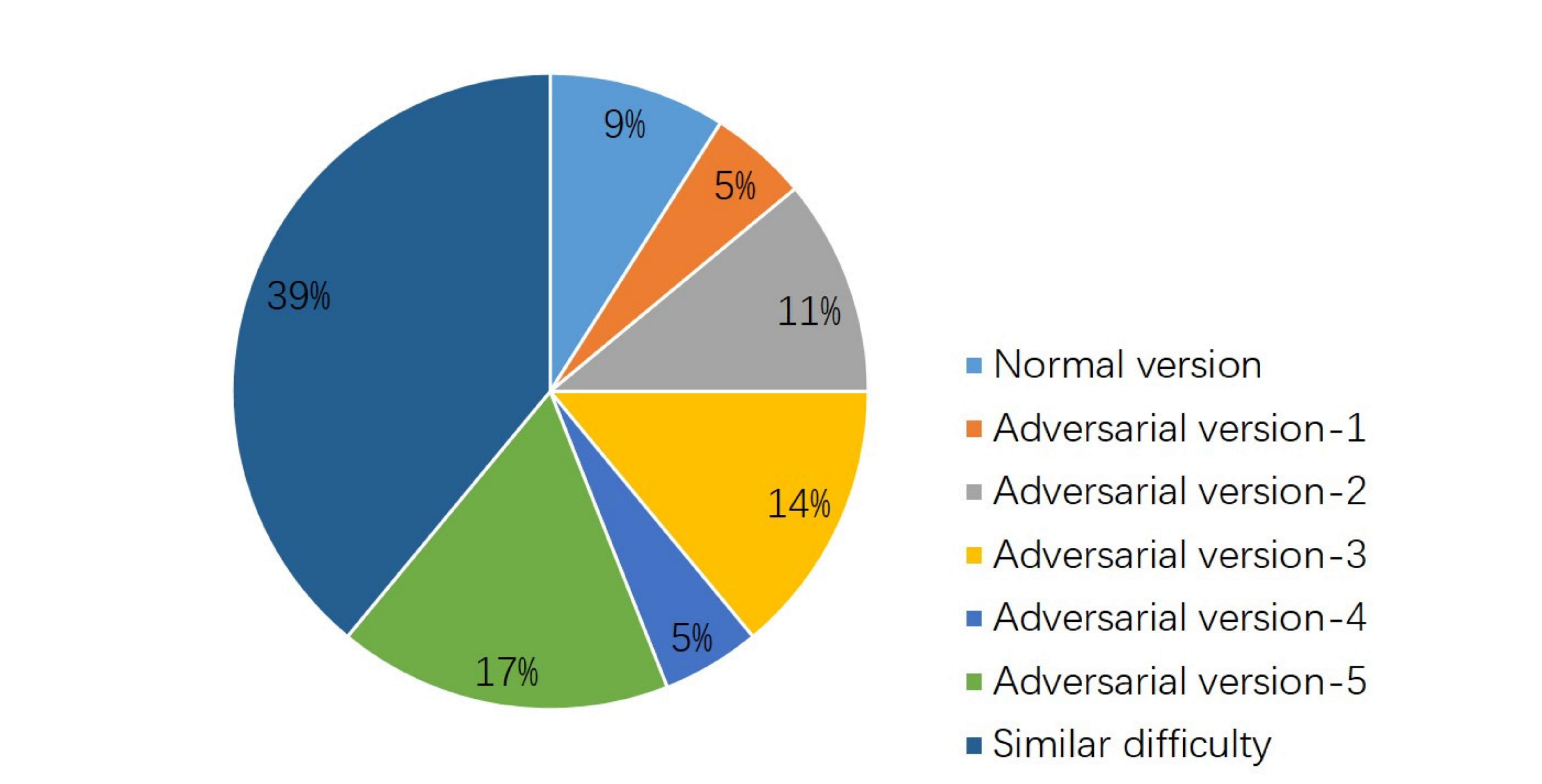}
	}
	\caption{Difficulty of normal and adversarial CAPTCHAs.} \label{f_difficulty}
\end{figure}

\subsection{Further Analysis}

Following the evaluation procedure, we ask some feedbacks
of users after finishing the CAPTCHA recognition tasks.
The first question is that \emph{which CAPTCHA is the most
	difficult one for recognition?}
The results are shown in Fig.\ref{f_difficulty}.
\begin{itemize}
	\item
	From Fig.\ref{f_difficulty} (a), in the text-based context, we can find that
	$67\%$ users think that adversarial and normal CAPTCHAs
	have similar difficulty, $19\%$ users think that
	adversarial CAPTCHAs are more difficult for recognition,
	and interestingly, there are also $14\%$ users think that the
	normal versions are more difficult.
	This indicates that adversarial CAPTCHAs do not increase the
	recognition difficulty obviously from the view of users.
	
	\item
	From Fig.\ref{f_difficulty} (b), in the image-based context,
	we can find that the users that think adversarial and normal CAPTCHAs
	have similar difficulty take the largest portion, saying $39\%$,
	while the other six options are varied from $5\%$ to $17\%$.
	Still, no adversarial CAPTCHAs are obviously difficult than
	the normal versions. This again indicates that image-based adversarial
	and normal CAPTCHAs have similar difficulty.
\end{itemize}


\begin{figure} [!tp]
	\centering
	\includegraphics[width=2.5in]{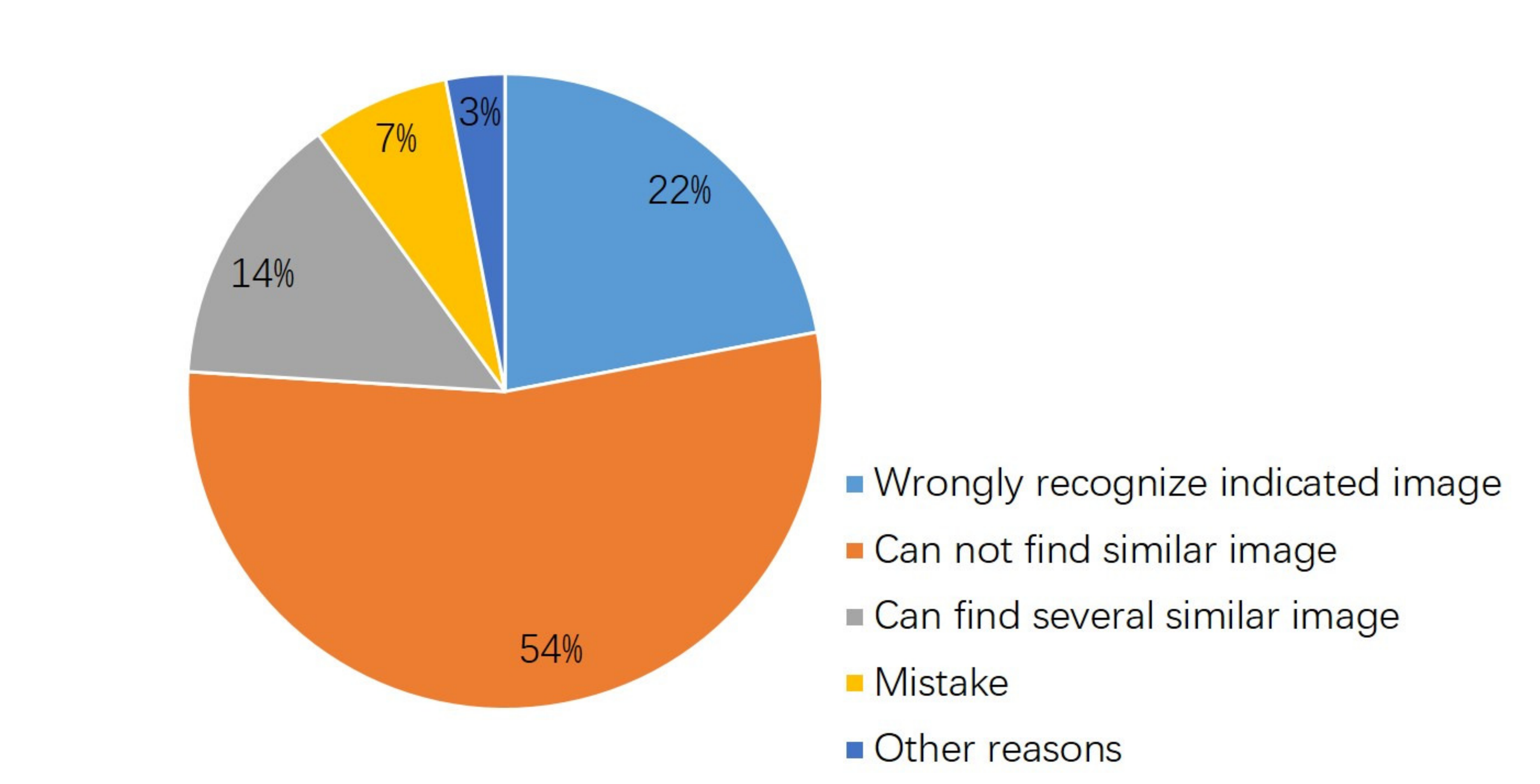}
	\caption{Reasons of failure recognition.} \label{f_failure}
\end{figure}

In Step 6 of the evaluation, if a user has one or more failures
in Steps 2-5, we will show her/him the failed tasks and ask a
question ``\emph{what is the most possible reason for this failure?}"
for each failed task.
We also provide five choices for this question: \emph{incorrectly recognize
	the source image}, \emph{can not find the target image}, \emph{find more than one target images}, \emph{mistakes},
and \emph{other reasons}.
After analyzing the collected data, we find that $24\%$ users
successfully finished all the CAPTCHA recognition tasks without any failure.
For the rest of the users, their feedbacks are shown in Fig.\ref{f_failure}.

From Fig.\ref{f_failure}, we can find that most of the failures
are caused by either cannot recognize the source image
or cannot recognize the target image.
We conjecture the main reason is that some of the randomly
selected images from ILSVRC-2012 might be semantically improper,
which are difficult to understand their semantical meanings
and further distinguish them.
Furthermore, most of the users finish the evaluation on their
smart phones. The relatively small screens may harm the recognizability
of images.

\section{Discussion} \label{limitation}


\noindent\textbf{Remarks on aCAPTCHA.}
Different from traditional CAPTCHA designs, which are mainly focusing on
defending against attacks in a \emph{passive} manner,
we design aCAPTCHA following a more proactive principle:
\emph{to set one's own spear against one's own shield}.
Then, in terms of the model of state-of-the-art CAPTCHA attacks,
we designed and implemented text- and image-based adversarial
CAPTCHAs.

When implementing adversarial CAPTCHAs, we also follow a different
methodology from that of existing adversarial image generation techniques.
The main reason, as we discussed before, is because we stand on a different
position. Existing adversarial image generation techniques focus on attacks
in a hidden manner. For instance, some method may focus on generating
an adversarial image which is only different from the original image
in one pixel \cite{suvararxiv17} (it is impossible for humans
to identify such difference).
In contrast, we follow the rule to inject as much perturbation as possible
when the adversarial CAPTCHAs remain human-tolerable.
By this way, we would find a better balance between CAPTCHA security
and usability, which can be demonstrated by our evaluation results.


One thing deserves further emphasis is that: \emph{aCAPTCHA is not designed
	as a replacement while is designed as an enhancement of existing CAPTCHA systems.}
According to our design, aCAPTCHA can be seamlessly combined
with the deployed text- and image-based CAPTCHA systems.
The only change is to update the normal CAPTCHAs with their
adversarial versions.
Therefore, we believe aCAPTCHA has a great applicability.
Actually, we have contacted with several Internet companies to introduce
aCAPTCHA. They are all very interested with aCAPTCHA and
two of them have shown the intension to integrate aCAPTCHA to their systems.

Finally, we believe open source is an important way to promote
computer science research, especially in the CAPTCHA defense domain.
Therefore, we make the aCAPTCHA system publicly available
at \cite{acaptcha}, including the source code, trained models,
datasets, as well as the usability evaluation interfaces.

\textbf{Limitations and Future Work.}
As an attempt to design adversarial CAPTCHAs, we believe aCAPTCHA can be
improved in many perspectives. We discuss the limitations of this work
along with future work below.

First, in the design of aCAPTCHA, we only integrate the popular
attacks to text- and image-based CAPTCHAs.
Also, following our design principle, we propose and implement
four text-based and four image-based adversarial CAPTCHA generation
methods, respectively.
Note that, all these designs and implementations are for demonstrating
the advantages of adversarial CAPTCHAs.
Furthermore, aCAPTCHA employs a modular design style, which is easy
for new technique integration.
Hence, we will add more attacks as well as more adversarial CAPTCHA
generation methods to aCAPTCHA, especially the emerging techniques.
We believe the open source nature will facilitate the improvement
process of aCAPTCHA.

Second, as we discussed, adversarial CAPTCHAs expect human-tolerable
instead of human-imperceptible perturbations.
However, in our evaluation, we set the human-tolerable perturbation
based on our experience and preliminary evaluation in our experiments,
i.e., we do not have a standard to quantify human-tolerable
perturbation yet. Therefore, it is expected to conduct more dedicated
research in understanding and quantifying the tradeoff between
CAPTCHA security and usability.

Third, in the paper, we do not consider that CAPTCHAs were being outsourced
to human labor. By design, CAPTCHAs are simple and easy to solve by humans 
while hard to solve by automated bot. This quality has made them easy to 
outsource to the global unskilled labor market. This type of attack is
hard to prevent. The function of CAPTCHAs is only to distinguish between 
the machine and the human. We should to design complementary
system to against human labor attack. This task is another interesting
future research topic.



\section{Related Work} \label{related}


\subsection{Traditional CATPCHAs}

\textbf{Text-based CAPTCHAs.}
The robustness of text-based CAPTCHAs is always an active research field.
In \cite{chesimnips05}, Chellapilla and Simard studied the security
of early text-based CAPTCHAs and proposed an effective machine learning
based attack to break them.
In \cite{burmarccs11}, Bursztein et al. conducted a systematic study
on the security of text-based CAPTCHAs with anti-segmentation techniques.
In \cite{yanahmccs08}, Yan and Ahmad found that the
\emph{Crowding Characters Together} (CCT) mechanism
could improve the security of CAPTCHAs.
However, such kind of security mechanisms are broken soon
by a group of attacks that leverage better machine learning
techniques \cite{ahmyantf11}\cite{buraigwoot14}.
Recently, Gao et al. demonstrated another simple yet powerful
machine learning based
attack that can break a wide range of text-based CAPTCHAs.
In a word, \emph{text-based CAPTCHA attacks continuously emerging
	while the defense research is far from enough}.

\noindent\textbf{Image-based CAPTCHAs.}
As another popular topic, image-based CAPTCHAs also draw a lot of
attention \cite{datliicm05}\cite{ruiliums04}\cite{elsdouccs07}.
In \cite{chetygicis04}, Chew and Tygar proposed three image-based
CAPTCHA schemes, which are still in wide use now.
On the other hand, in \cite{golccs08}, Golle developed a machine learning
based attack against the Asirra CAPTCHA.
Moreover, in \cite{zhuyanccs10}, Zhu et al. systematically studied
the design of image-based CAPTCHAs and showed an attack to
break 12 existing CAPTCHA schemes.
Following another track,  Sivakorn et al. designed a novel attack
that leverages online image annotation services and libraries \cite{sivpoleurosp16}.
Similar to the text-based CAPTCHA scenario, \emph{more defensive research
	is also expected to secure image-based CAPTCHAs}.

\noindent\textbf{Other CAPTCHAs.}
There are also many other forms of CAPTCHAs, such as audio-based CAPTCHAs
\cite{gaoliuisecs10}, video-based CAPTCHAs \cite{kluzansoups09},
game-based CAPTCHAs \cite{mohsaciccs14}, etc.
However, those CAPTCHAs are not widely employed in practice mainly
because of the usability issue.
Furthermore, there also exist plenty of attacks that can break them
\cite{bocpatwoot17}\cite{mohgaocs17}\cite{xureyusenixsecurity12}.

\subsection{Emerging CAPTCHAs}


\noindent\textbf{DeepCAPTCHA. }
In \cite{osahertifs17}, Osadchy et al. introduced a new image-based CAPTCHA scheme 
which is designed to resist machine learning attacks. It adds \emph{Immutable Adversarial
Noise} (IAN) to the correctly classified images that deceive deep learning tools and cannot
be removed using image filtering. However,
DeepCAPTCHA is different from our approach. In general, DeepCAPTCHA is a new type of image-based
CAPTCHA scheme which could provide high security.   
While our aCAPTCHA system is designed to enhance the existing CAPTHCA schemes.
Furthermore, the proposed IAN, which is resistance to filtering attack, cannot be used in
text-based CAPTCHA generation.
In this work, we consider more state-of-the-art adversarial example defense strategies and propose several new
methods to generate text- or image- based adversarial CAPTCHAs.

\noindent\textbf{reCAPTCHA. }
The reCAPTCHA service offered by Google is the most widely used CAPTCHA service.
It is a new multi-stage CAPTCHA system\cite{sivpoleurosp16}. At the first round of check-authentication,
Google leverages information about user's activities to correlate requests to 
users that have previously interacted with any of its services. 
If the user is deemed legitimate, he is not required to solve a challenge. Otherwise,
the user needs to further solves the given text- or image- based CAPTCHA correctly.
reCAPTCHA and aCATPCHA do not conflict. aCAPTCHA can be used for
further improving reCAPTCHA's security.

\subsection{Defense Methods against Adversarial Examples}
The robustness of machine learning models against adversarial examples is an
active research filed recently.
In \cite{szezararxiv13}, Szegedy et al. found that adversarial training increases the
robustness of a model by augmenting training data with adversarial examples.
In \cite{madmakarxiv17}, Madry et al. showed that adversarially trained models can be
more robust against white-box attacks if the perturbation during training closely
maximizes the model's loss.
In \cite{trakuriclr2018} , Tramer et al. propsed ensemble adversarial training,
a technique that augments training data with perturbation transferred from other
models. It can somehow make a model resist to black-box attacks.

Another way to defend against adversarial perturbation is input transformation.
Without of changing the model structure, it tries to eliminate the perturbation in the input.
Xu et al.\cite{xuevandss18} proposed feature squeezing, reducing the color bit depth and spatial 
smoothing. These simple strategies are inexpensive and can sever as complementary to other
defenses. In \cite{guoraniclr18}, Guo et al. ensembled various input transformations to
counter adversarial images. However, these methods are not strong when against white-box attacks,
and can be broken by minor modifications \cite{athcarICML2018}. Moreover, Athalye et al. 
\cite{athcarICML2018} described that gradient masking is an incomplete defense to adversarial examples. Many state-of-the-art
gradient masking schemes can be successfully circumvented by their attacks. 


\section{Conclusion} \label{conclusion}

In this paper, we study the generation of adversarial CAPTCHAs.
First,
we propose a framework for generating text-
and image-based adversarial CAPTCHAs.
Then, we design and implement aCAPTCHA, a comprehensive
adversarial CAPTCHA generation and evaluation system,
which integrates 10 image preprocessing techniques,
9 CAPTCHA attacks, 4 baseline adversarial CAPTCHA generation methods,
and 8 new adversarial CAPTCHA generation methods,
and can be used for the generation, security evaluation,
and usability evaluation of adversarial CAPTCHAs.
To evaluate the performance of aCAPTCHA, we conduct
extensive experiments. The results demonstrate that
the adversarial CAPTCHAs generated by aCAPTCHA can significantly
improve the security of normal CAPTCHAs while maintaining similar usability.
Finally, we open source aCAPTCHA to facilitate the
CAPTCHA security research.

%
%

\ifCLASSOPTIONcaptionsoff
  \newpage
\fi



%




\end{document}